\documentclass[preprint,5p,twocolumn,number,11pt]{elsarticle}

\usepackage{graphics}

\usepackage{amssymb}
\usepackage{amsthm}

\usepackage{ulem}
\usepackage{tabularx}
\usepackage{textcomp}
\usepackage[english]{babel}

\journal{Astroparticle Physics}

\begin{document}

\begin{frontmatter}


\title{Comparing LOPES measurements of air-shower radio emission with REAS 3.11 and CoREAS simulations\\ \small{\textbf{Results partly outdated due to new calibration:} see Astroparticle Physics 75 (2016) 72-74 (arxiv:1507.07389).}}

\author[1]{W.D.~Apel}
\author[14]{J.C.~Arteaga-Vel\'azquez}
\author[3]{L.~B\"ahren}
\author[1]{K.~Bekk}
\author[4]{M.~Bertaina}
\author[5]{P.L.~Biermann}
\author[1,2]{J.~Bl\"umer}
\author[1]{H.~Bozdog}
\author[6]{I.M.~Brancus}
\author[4,8]{E.~Cantoni}
\author[4]{A.~Chiavassa}
\author[1]{K.~Daumiller}
\author[15]{V.~de~Souza}
\author[4]{F.~Di~Pierro}
\author[1]{P.~Doll}
\author[1]{R.~Engel}
\author[3,9,5]{H.~Falcke}
\author[2]{B.~Fuchs}
\author[10]{D.~Fuhrmann}
\author[11]{H.~Gemmeke}
\author[7]{C.~Grupen}
\author[1]{A.~Haungs}
\author[1]{D.~Heck}
\author[3]{J.R.~H\"orandel}
\author[5]{A.~Horneffer}
\author[2]{D.~Huber}
\author[1]{T.~Huege}
\author[16]{P.G.~Isar}
\author[10]{K.-H.~Kampert}
\author[2]{D.~Kang}
\author[11]{O.~Kr\"omer}
\author[3]{J.~Kuijpers}
\author[2]{K.~Link}
\author[12]{P.~{\L}uczak}
\author[2]{M.~Ludwig}
\author[1]{H.J.~Mathes}
\author[2]{M.~Melissas}
\author[8]{C.~Morello}
\author[1]{J.~Oehlschl\"ager}
\author[2]{N.~Palmieri}
\author[1]{T.~Pierog}
\author[10]{J.~Rautenberg}
\author[1]{H.~Rebel}
\author[1]{M.~Roth}
\author[11]{C.~R\"uhle}
\author[6]{A.~Saftoiu}
\author[1]{H.~Schieler}
\author[11]{A.~Schmidt}
\author[1]{F.G.~Schr\"oder\corref{cor}}
\ead{frank.schroeder@kit.edu}
\author[13]{O.~Sima}
\author[6]{G.~Toma}
\author[8]{G.C.~Trinchero}
\author[1]{A.~Weindl}
\author[1]{J.~Wochele}
\author[12]{J.~Zabierowski}
\author[5]{J.A.~Zensus}

\address[1]{Institut f\"ur Kernphysik, Karlsruhe Institute of Technology (KIT), Germany}
\address[{14}]{Universidad Michoacana, Morelia, Mexico}
\address[3]{Radboud University Nijmegen, Department of Astrophysics, The Netherlands}
\address[4]{Dipartimento di Fisica Generale dell' Universit\`a Torino, Italy}
\address[5]{Max-Planck-Institut f\"ur Radioastronomie Bonn, Germany}
\address[2]{Institut f\"ur Experimentelle Kernphysik, Karlsruhe Institute of Technology (KIT), Germany}
\address[6]{National Institute of Physics and Nuclear Engineering, Bucharest, Romania}
\address[8]{INAF Torino, Instituto di Fisica dello Spazio Interplanetario, Italy}
\address[{15}]{Universidad S\~ao Paulo, Inst. de F\'{\i}sica de S\~ao Carlos, Brasil}
\address[9]{ASTRON, Dwingeloo, The Netherlands}
\address[{10}]{Universit\"at Wuppertal, Fachbereich Physik, Germany}
\address[{11}]{Institut f\"ur Prozessdatenverarbeitung und Elektronik, Karlsruhe Institute of Technology (KIT), Germany}
\address[7]{Universit\"at Siegen, Fachbereich Physik, Germany}
\address[{16}]{Institute of Space Science, Bucharest, Romania}
\address[{12}]{National Centre for Nuclear Research, Department of Cosmic Ray Physics, {\L}\'{o}d\'{z}, Poland}
\address[{13}]{University of Bucharest, Department of Physics, Romania}

\cortext[cor]{Corresponding author:}

\begin{abstract}
Cosmic ray air showers emit radio pulses at MHz frequencies, which can be measured with radio antenna arrays -- like LOPES at the Karlsruhe Institute of Technology in Germany. To improve the understanding of the radio emission, we test theoretical descriptions with measured data. The observables used for these tests are the absolute amplitude of the radio signal, and the shape of the radio lateral distribution. We compare lateral distributions of more than 500 LOPES events with two recent and public Monte Carlo simulation codes, REAS 3.11 and CoREAS (v 1.0). The absolute radio amplitudes predicted by REAS 3.11 are in good agreement with the LOPES measurements. The amplitudes predicted by CoREAS are lower by a factor of two, and marginally compatible with the LOPES measurements within the systematic scale uncertainties. In contrast to any previous versions of REAS, REAS 3.11 and CoREAS now reproduce the shape of the measured lateral distributions correctly. This reflects a remarkable progress compared to the 
situation a few years ago, and it seems that the main processes for the radio emission of air showers are now understood: The emission is mainly due to the geomagnetic deflection of the electrons and positrons in the shower. Less important but not negligible is the Askaryan effect (net charge variation). Moreover, we confirm that the refractive index of the air plays an important role, since it changes the coherence conditions for the emission: Only the new simulations including the refractive index can reproduce rising lateral distributions which we observe in a few LOPES events. Finally, we show that the lateral distribution is sensitive to the energy and the mass of the primary cosmic ray particles.
\end{abstract}

\begin{keyword}
cosmic rays \sep extensive air showers \sep radio emission \sep LOPES \sep lateral distribution
\end{keyword}

\end{frontmatter}


\section{Introduction}

Hundred years after the discovery of cosmic rays \cite{Hess1912} the origin of the highest energy particles is still unclear and requires further measurements. The interesting energy range above $10^{17}\,$eV where the transition from galactic to extragalactic cosmic rays is presumed \cite{Haungs2003} cannot be accessed by direct measurements since the flux of cosmic rays above $\sim 10^{14}\,$eV is too low. Instead, cosmic rays are measured indirectly by detecting air showers of secondary particles. Established techniques for air shower measurements are the detection of the secondary particles at ground, and the measurement of atmospheric Cherenkov and fluorescence light emitted by air showers. The aim of all measurements is to reconstruct the properties of the primary cosmic ray particles from the air-shower observables, i.e.~their arrival direction, energy and mass. In particular, the latter two techniques give more precise measurements of the energy since they provide a calorimetric measurement of the 
air showers. However, they have the disadvantage that they can be used only during dark, moonless nights \cite{Bluemer2009}.

An alternative instrument for air shower detection is given by digital radio antenna arrays, which also can provide a measurement of the shower energy \cite{HuegeUlrichEngel2008, PalmieriICRC2013} and feature a duty-cycle of almost $100\,\%$ like particle detector arrays \cite{Allan1971, Melissas2010}. Current efforts like LOPES \cite{FalckeNature2005, HuegeARENA2010, SchroederLOPES_ARENA2012}, CODALEMA \cite{ArdouinBelletoileCharrier2005, RavelARENA2010}, at ANITA \cite{Hoover2010ANITA}, the Pierre Auger Observatory \cite{FuchsRICAP2011}, or at Tunka \cite{SchroederTunkaRex_ARENA2012} still focus on engineering work, i.e.~to prove the applicability of the radio technique to large scale observatories, and to show that a precision similar to the one of the fluorescence and air-Cherenkov techniques can be achieved.

To make the radio technique competitive, it is crucial to understand the radio emission mechanism in sufficient detail. From previous work (e.g., Ref.~\cite{Allan1971}) we know that the radio emission of air showers originates mainly from the geomagnetic deflection of electrons and positrons \cite{KahnLerche1966, FalckeGorham2003}, but also other effects play a role, in particular the Askaryan effect \cite{Askaryan1962}, i.e.~the variation of the net charge excess over the shower development. Recent models and simulation codes also take into account the refractive index of the air, which affects the coherence conditions for all radio emission mechanisms \cite{deVries2011, LudwigICRC2011, AlvarezZHAires2012}, as was already discussed more than 40 years ago in Ref.~\cite{AllanICRC1971}. Cherenkov emission due to constant, not-accelerated charges moving with superluminal velocity in a refractive medium, however, is generally not included in recent simulation codes because its contribution is considered 
negligible \cite{James2011}. The best way to test our understanding of the overall emission is to compare experimentally measured quantities, like the lateral distribution of the radio signal, with simulations based on certain models. 

The lateral distribution of the radio signal is the variation of the radio amplitude $\epsilon$ with the distance to air shower axis $d$. It has already been studied with LOPES \cite{ApelArteagaAsch2010, ApelLOPES_MTD2012} and other experiments (e.g., CODALEMA \cite{RavelARENA2010}), and already it was compared to simulations \cite{SchroederECRS2010}, however, only for single events or with a limited statistics. These limited studies had been sufficient to show that earlier models, not yet including the refractive index of the air, could not fully reproduce the measurements. Now, we present the results of a systematic, per-event comparison of radio lateral distributions measured with LOPES to two recent, publicly available Monte Carlo simulation codes: REAS 3.11 \cite{LudwigHuege2010} and CoREAS (v 1.0) \cite{HuegeCoREAS_ARENA2012}, respectively. Since LOPES features an absolute amplitude calibration \cite{NehlsHakenjosArts2007}, and neither REAS nor CoREAS have free parameters to tune the absolute scale, 
the presented comparison is not only qualitatively, but also quantitatively meaningful. 

There are other simulation codes available, e.g., ZHAireS \cite{AlvarezZHAires2012}, EVA \cite{Werner2012}, MGMR \cite{ScholtenMGMR2008}, SELFAS \cite{Marin2012} and the model described in Ref.~\cite{KalmykovKonstantinov2011}. Although they use different approaches and techniques, they basically simulate the same physics for the radio emission of air showers, i.e., the geomagnetic and the Askaryan effect, and some codes also include the refractive index of the air. Since we have chosen on purpose two simulation codes with a fixed, non-tunable absolute scale, other codes can be indirectly compared to LOPES when comparing them with REAS 3.11 or CoREAS, respectively. Such model-to-model comparisons have already been performed, e.g., in Refs.~\cite{HuegeARENAmodels2010, HuegeARENAtheory2012}. In future, the next-generations experiments LOFAR \cite{BuitinkLOFAR_ICRC2013} and AERA \cite{SchroederAERA_ICRC2013} can be used to test simulations in much more detail or, respectively, at much larger axis distances.

\section{Measurements}
For the reconstruction of the lateral distribution we use LOPES measurements of the years 2005-2009. During this period, LOPES consisted of 30 absolute calibrated, inverted v-shaped dipole antennas. Until the end of 2006, all antennas were east-west aligned, since for most shower geometries the radio signal is preferentially east-west polarized due to the geomagnetic radio emission. Afterwards, half of the antennas were aligned in the north-south direction. In this paper we focus on the lateral distribution measured with the east-west aligned antennas, since the statistics are larger and the signal-to-noise ratio is generally higher. At the end of the paper, we also present first results for lateral distributions measured with the north-south aligned antennas.

The LOPES antennas are located within the KASCADE-Grande experiment \cite{ApelArteagaBadea2010} at the Karlsruhe Institute of Technology, Germany, and form an array with a lateral extension of about $200\,$m. Whenever KASCADE-Grande detects a high-energy shower ($10^{16}-10^{18}\,$eV), it triggers LOPES which then digitally measures the radio emission. The effective bandwidth of LOPES is $43-74\,$MHz, the trace length about $0.8\,$ms, and the sampling rate $80\,$MHz. This means that the LOPES setup fulfills the Nyquist sampling theorem, i.e.~the full information of the electrical field strength in the measurement bandwidth is contained in the data. Thus, the electrical field strength between the sampled data points is retrieved by up-sampling.

\begin{figure*}[t]
  \centering
  \includegraphics[width=0.92\columnwidth]{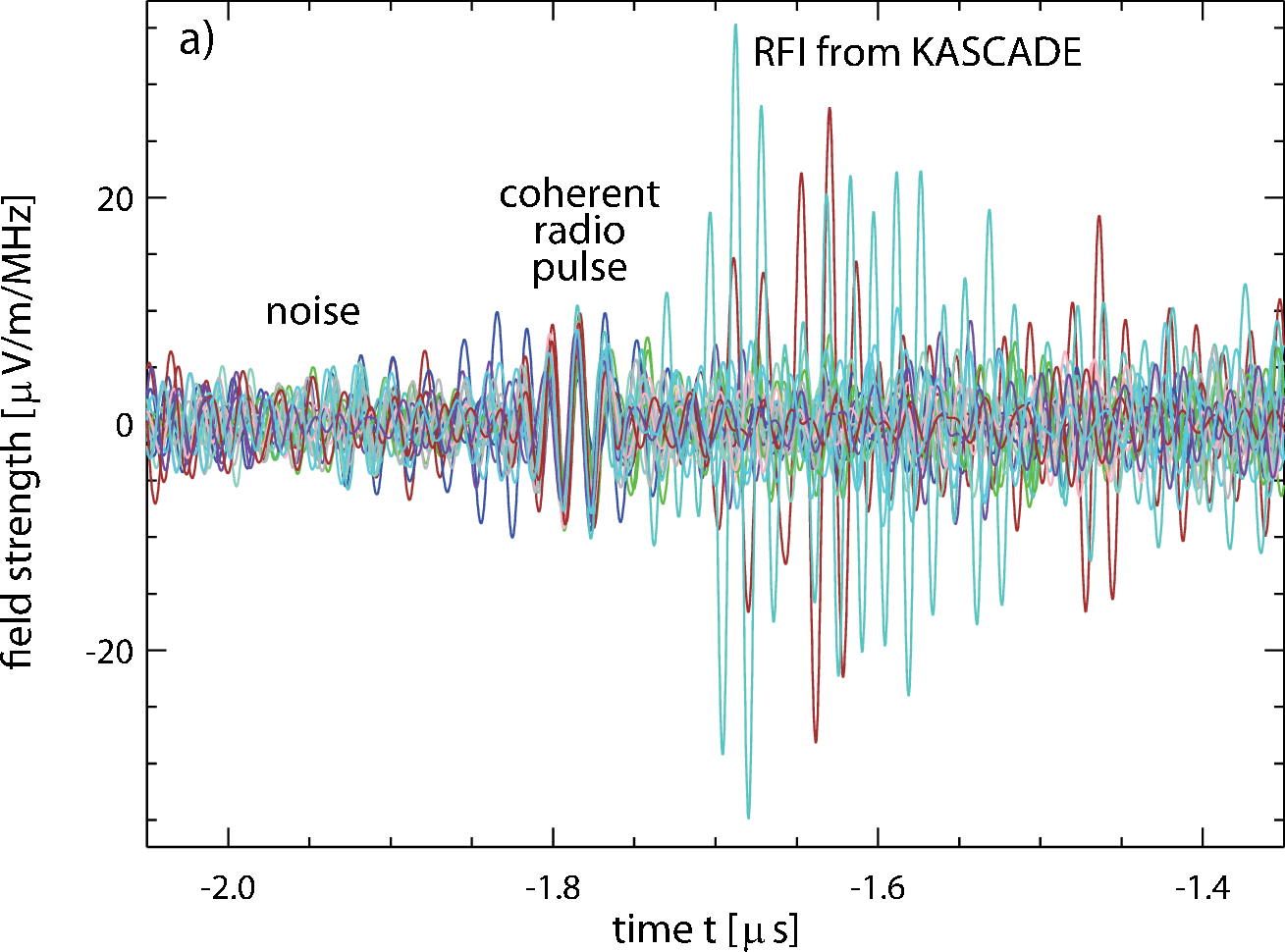}
  \hskip 0.13\columnwidth
  \includegraphics[width=0.92\columnwidth]{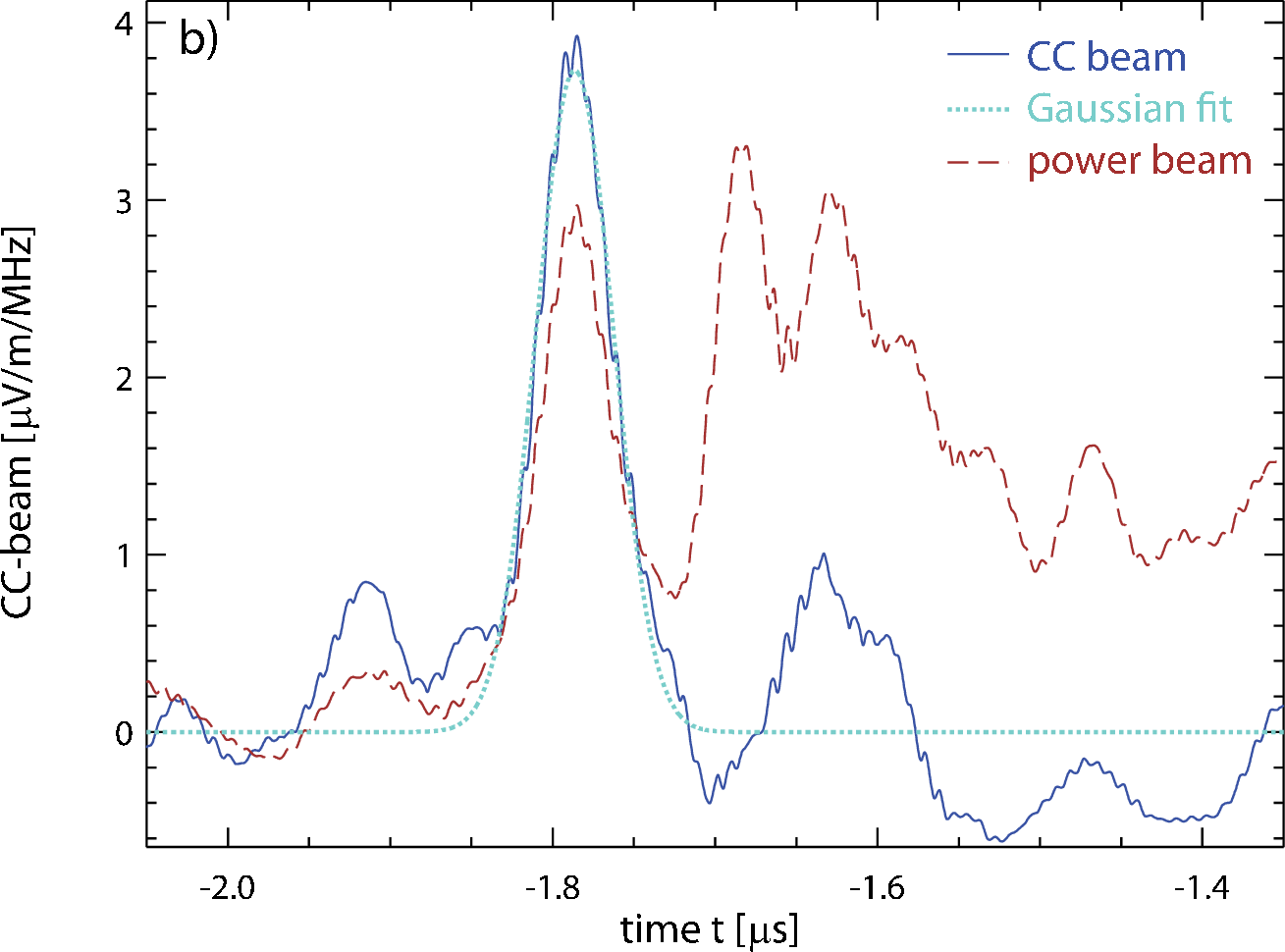}
  \vskip 0.2 cm
  \includegraphics[width=0.92\columnwidth]{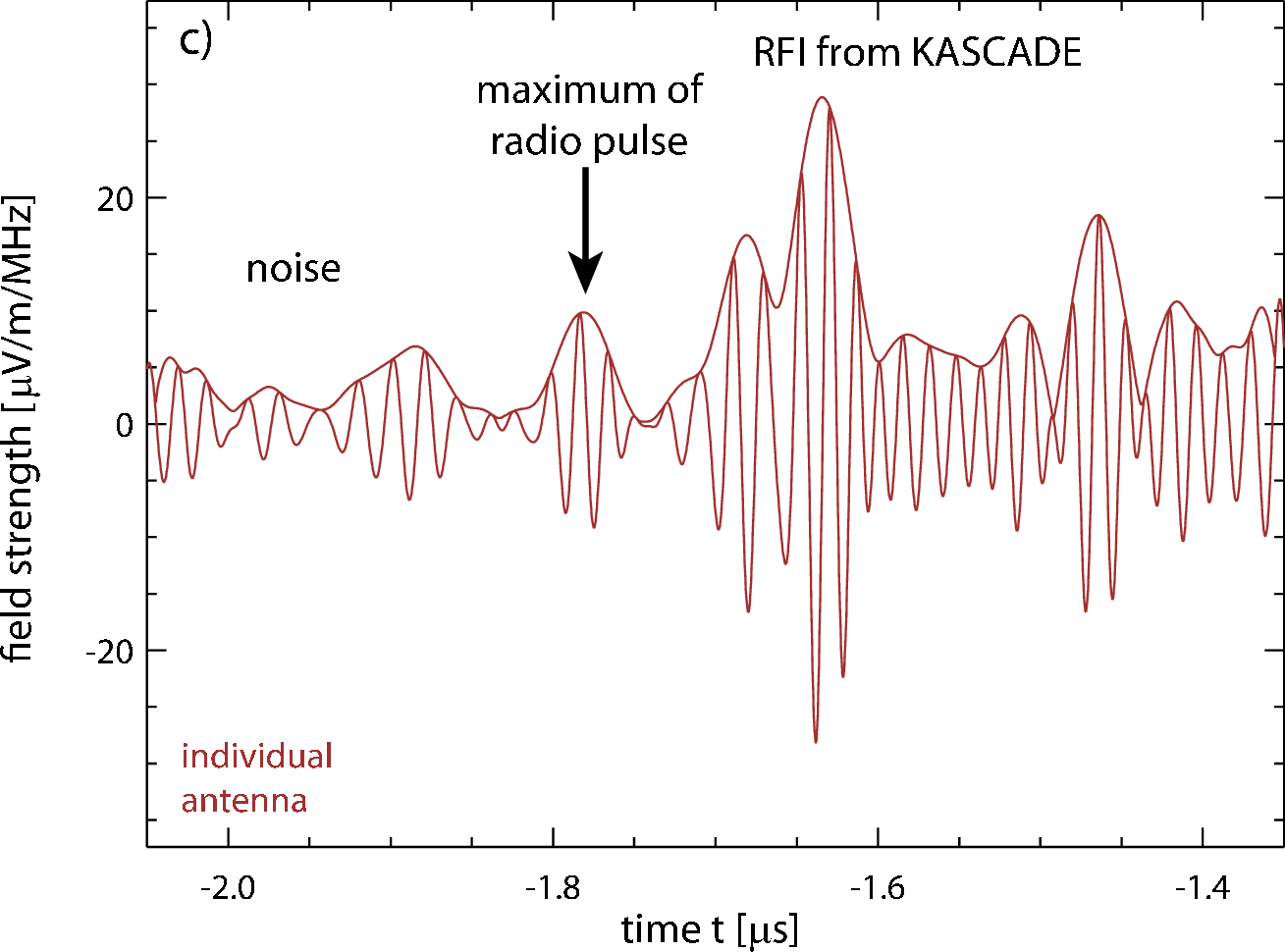}
  \hskip 0.13\columnwidth
  \includegraphics[width=0.92\columnwidth]{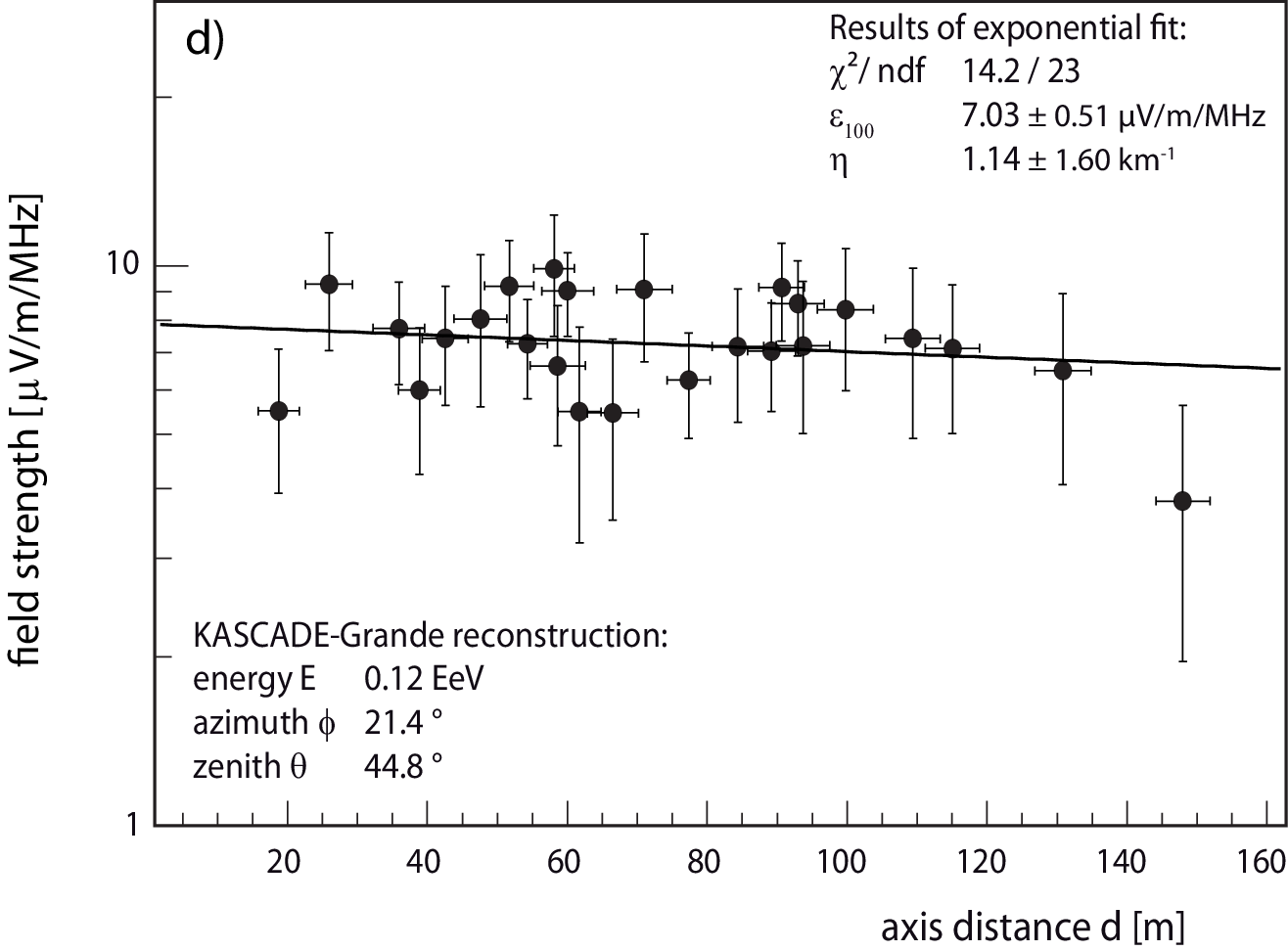}
  \caption{Example event measured with LOPES in 2005, when all antennas were aligned in the east-west direction: electrical field strength traces of all antennas (a), smoothed cross-correlation (CC) beam and power beam (b), electrical field strength trace in an individual antenna with a Hilbert envelope used to measure the maximum amplitude (c), lateral distribution (d).}
   \label{fig_exampleEvent}
 \end{figure*}

\subsection{Analysis Procedure}

The KASCADE-Grande reconstruction provides the shower direction and the shower core (= intersection of the air shower axis with the ground), as well as the primary energy, and the measured numbers of muons and electromagnetic particles. A subset of these parameters (arrival direction, core position, energy, muon number) is used as input for the REAS 3.11 and CoREAS simulations, and another subset (shower direction, core position) as input for the LOPES analysis, whereby the shower direction is optimized during the LOPES analysis using the radio measurements.\footnote{The LOPES analysis software is available as open-source at \lq http://usg.lofar.org\rq.}

An essential part of the LOPES analysis is cross-correlation beamforming. First, the electrical field strength traces of each antenna are shifted in time corresponding to the arrival direction of the air shower (Fig.~\ref{fig_exampleEvent} a), and, second, a cross-correlation between the different traces is calculated (CC beam, Fig.~\ref{fig_exampleEvent} b) as well as a power beam measuring the total power from the air shower arrival direction. Details of this procedure can be found in Refs.~\cite{HornefferThesis2006, HornefferICRC2007}. The present analysis of the lateral distribution uses the cross-correlation and power beam only for two purposes: a selection of events with a clear radio signal, and a determination of the exact time of the radio pulse, which is an important input to determine the radio amplitude in antennas with low signal-to-noise ratio.

The radio signal in each individual antenna is determined with a Hilbert envelope of the up-sampled trace (Fig.~\ref{fig_exampleEvent} c), where the maximum of the Hilbert envelope is a measure for the maximum instantaneous amplitude of the radio signal. Since the time of the radio pulse is known from the preceding cross-correlation beamforming, a measurement of the radio amplitude at this time is possible in each individual antenna. For this, we take the maximum of the Hilbert envelope closest to the time of the CC beam maximum as input for the lateral distribution (Fig.~\ref{fig_exampleEvent} d). To convert the measured radio pulse amplitudes into the absolute electrical field strength, we use simulations of our antenna gain pattern and calibration measurements of the complete electronics chain including the antenna. However, for a correct conversion of the measured amplitudes to the east-west or, respectively, the north-south polarization component of the field strength vector, a measurement with at least 
two differently aligned antennas at the same location would be required. Since this is only available for about half of the LOPES events and a few antenna positions, we generally use a simplified conversion formula, which is exact in the limit that the radio signal is purely east-west polarized. For the simulations, we determine the pulse amplitude by digitally filtering the east-west polarization component of the simulated radio signal to the effective bandwidth of LOPES (cf. section \ref{sec_simulations}).

We point out that a different analysis procedure (e.g., using the integrated pulse power instead of the amplitude) might result in a different lateral distribution. The lateral distribution likely depends on the used frequency band, too. Theoretical models predict that the pulse shape and thus also the shape of the frequency spectrum of the radio emission depends on the distance to the air shower axis \cite{HuegeARENAmodels2010}. A direct test of our results will still be possible when results of measurements with a larger effective bandwidth are digitally filtered to the effective bandwidth of LOPES ($43-74\,$MHz). To facilitate a comparison with other experiments, we have normalized the electrical field strength dividing by the effective bandwidth of LOPES, though this will allow only a comparison at first order, since the frequency spectrum is not flat \cite{NiglFrequencySpectrum2008, GrebeSpectralSlope_ARENA2012}.

\begin{figure}
\centering
\includegraphics[width=0.48\textwidth]{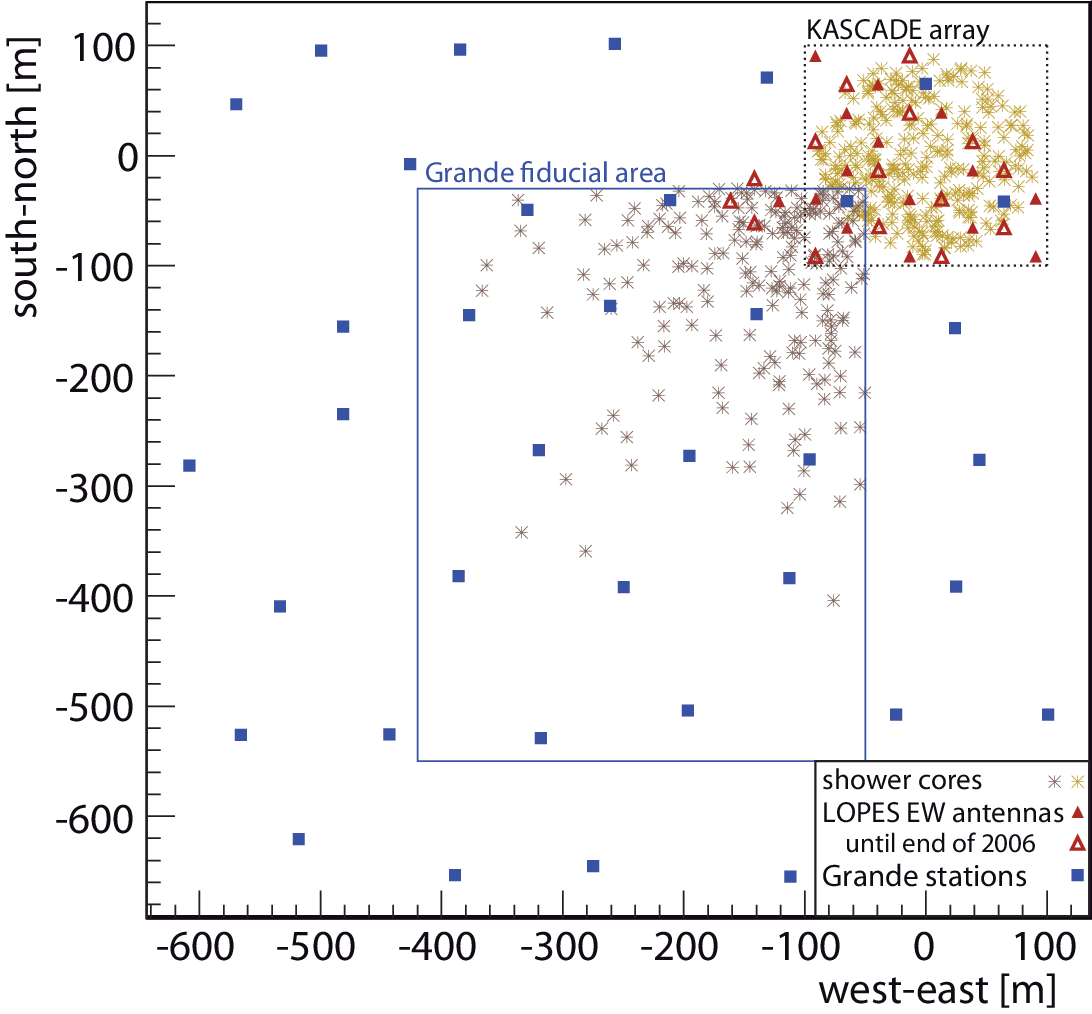}
\caption{Map of the LOPES array co-located with the KASCADE-Grande experiment, and shower cores of the selected LOPES events. Due to quality cuts of the KASCADE-Grande reconstruction, only events with a core location within the fiducial area of the KASCADE array or the Grande extension are accepted which gives rise to the distinct shapes of the core locations.} \label{fig_eventMap}
\end{figure}

\subsection{Event selection}

For the present study, we preselected 3968 LOPES events using the KASCADE-Grande measurement of the same events, requiring a minimum reconstructed energy of $10^{17}\,$eV and a zenith angle $\theta \le 45^\circ$. Then, we selected only events which show a clear signal in the CC beam, i.e.~the fraction of correlated power in the antennas (= height of CC beam / height of power beam) must be larger than $80\,\%$. In addition, the signal-to-noise ratio of the CC beam must be larger than $14$ (normalized with a factor $\sqrt{N_\textrm{ant}/30}$, since not all 30 LOPES antennas are available in each event). The signal-to-noise ratio is calculated as the height of a Gaussian fit to the CC beam smoothed by block-averaging, divided by the RMS of the smoothed CC beam in a part of the trace before the radio pulse. Furthermore, we exclude events for which we measured a high atmospheric electrical field at ground ($E_\mathrm{atm} > 3000\,$V/m), since this can significantly affect the radio emission of air showers \cite{
BuitinkApelAsch2006, LopesThunderstormARS2011}.

This way we attained 528 events. From these events, we exclude in total 15 events for different reasons: in 3 cases there have been technical problems with one of the performed simulations, in one cases the fit to the measured lateral distribution did not converge for technical reasons in the fit algorithms, in 14 cases the fit failed because at least one antenna measured a too low amplitude. Thus, 513 events remain for the lateral distribution analysis of this paper (see Fig.~\ref{fig_eventMap}). This is only a fraction of the preselected events, since LOPES is not fully efficient at the chosen energy threshold. For example, the efficiency depends on the angle between the Earth's magnetic field and the air shower axis (= geomagnetic angle $\alpha$), since the signal height depends on the geomagnetic Lorentz force. Furthermore, preselected events closer to the LOPES antennas are more likely to pass the quality cuts than events at larger axis distances, since the radio amplitude decreases with increasing 
distance. However, since we have done the simulations for exactly the selected air showers, neither full efficiency nor a uniform distribution of the events is required to test the simulations with LOPES measurements.

\subsection{Lateral distribution function (LDF)}
According to experimental and theoretical studies, the 'true' lateral distribution function (LDF) is quite complex, cf.~Refs.~\cite{AllanICRC1971, deVries2010, FuchsPhD, NellesLOFAR_ARENA2012}: in the LOPES distance and frequency range, the LDF flattens towards the shower core, in some cases shows a bump at an axis distance of about $120\,$m and then shows an approximately exponential decrease. On top of that, the interference between the dominant geomagnetic and the weaker Askaryan effect leads to a small asymmetry depending on the observer azimuth relative to the shower core. According to Ref.~\cite{SchoorlemmerPhD}, the relative strength of the Askaryan effect compared to the geomagnetic effect is about $11.5\,\%$ at the location of the Pierre Auger Observatory, where the geomagnetic field is only half as strong as at the LOPES site. Thus, the relative size of this effect should be about $5-6\,\%$ at LOPES, which is small compared to the typical measurement uncertainty of the amplitude in an individual 
LOPES antenna. Consequently, the asymmetry of the lateral distribution is neglected when defining the LDF for the present analysis. Finding a function describing all these features correctly would be an interesting study on its own, but is not the focus of this paper.

Instead, we have decided to use a simplified 2-parameter function which still is sufficient to test whether the simulations can correctly describe the absolute amplitude scale and the general shape of the measured lateral distributions. Using a more complex LDF with additional parameters describing the features mentioned above would drastically reduce the statistics of usable events, because only for a small fraction of the events all fit parameters could be determined with sufficient precision. E.g., if there were one additional parameter to describe the flattening of the lateral distribution near the shower axis, the fit would be poorly confined for all events which have no antenna close to the shower axis. Thus, we have decided to use an exponential LDF with an amplitude and a slope parameter, as already in Refs.~\cite{ApelArteagaAsch2010, ApelLOPES_MTD2012} -- in particular since a power law as alternative 2-parameter LDF has been shown to be inappropriate for LOPES events \cite{ApelArteagaAsch2010}.

Even with the simple exponential LDF we can test whether the simulations reproduce the more complex features of the true lateral distribution in two different ways. First, we compare individual events and see generally a good agreement in the shape. Second, we compare the dependencies of the slope parameter on the shower inclination and the mean axis distance where the event is observed. The fact that there is a dependence reflects that the true LDF is indeed not a pure exponential. By comparing the measured dependence with the simulated one, we can consequently test (at least in first order) whether simulations would fail to reproduce the general shape of the true LDF. I.e., the complex nature of the true lateral distribution is reflected in the dependencies of the LDF on the shower parameters and can be tested this way. For this reason, testing simulations against measurements does not require a complex LDF describing all the detailed features of the lateral distribution. Instead, it is useful to choose a 
simple LDF which gives a sufficient description of each individual event. This allows for large statistics of events and a relatively low uncertainty of the fit parameters at the same time. For other purposes, the decision for a specific LDF might be different, though.

Following this goal, we slightly modified the exponential LDF used in Ref.~\cite{ApelArteagaAsch2010}: First, we use the amplitude at an axis distance of $100\,$m as fit parameter instead of the amplitude at the shower core, since $100\,$m is close to the mean axis distance of about $120\,$m and close to the distance where least fluctuations due to the primary particle type are expected for LOPES ($70 - 90\,$m) \cite{PalmieriLOPES_ARENA2012}. This reduces the uncertainty of the amplitude fit parameter while still describing the same physics. Second, we switched from the original slope parameter $R_0$ (distance over which the amplitude decreases by a factor 1/e) to its reciprocal $\eta=1/R_0$. There are two reasons for this change: First, some events show flat lateral distributions with an original slope parameter $R_0$ much larger than the lateral extension of LOPES, which is about $200\,$m. For such flat events, exact values of $R_0$ can differ a lot while the absolute values of $\eta$ are similar. Second, 
there are a few events with rising lateral distributions, i.e.~with a negative $\eta$, which for technical reasons would be difficult to fit using $R_0$. Thus, $\eta=1/R_0$ is the physically more sensible parameter. Therefore, we use the following fit function for the measured and simulated lateral distributions: 

\begin{equation}
\label{eq_ldfEXP}
\epsilon(d) = \epsilon_{100} \cdot \exp{(- \eta \cdot (d-100\,\mathrm{m}))} 
\end{equation}

with the amplitude (electrical field strength) $\epsilon(d)$ at a distance to the air shower axis $d$, the amplitude at $100\,$m as amplitude fit parameter $\epsilon_{100}$, and the slope fit parameter $\eta$.

\begin{figure}
\centering
\includegraphics[width=0.48\textwidth]{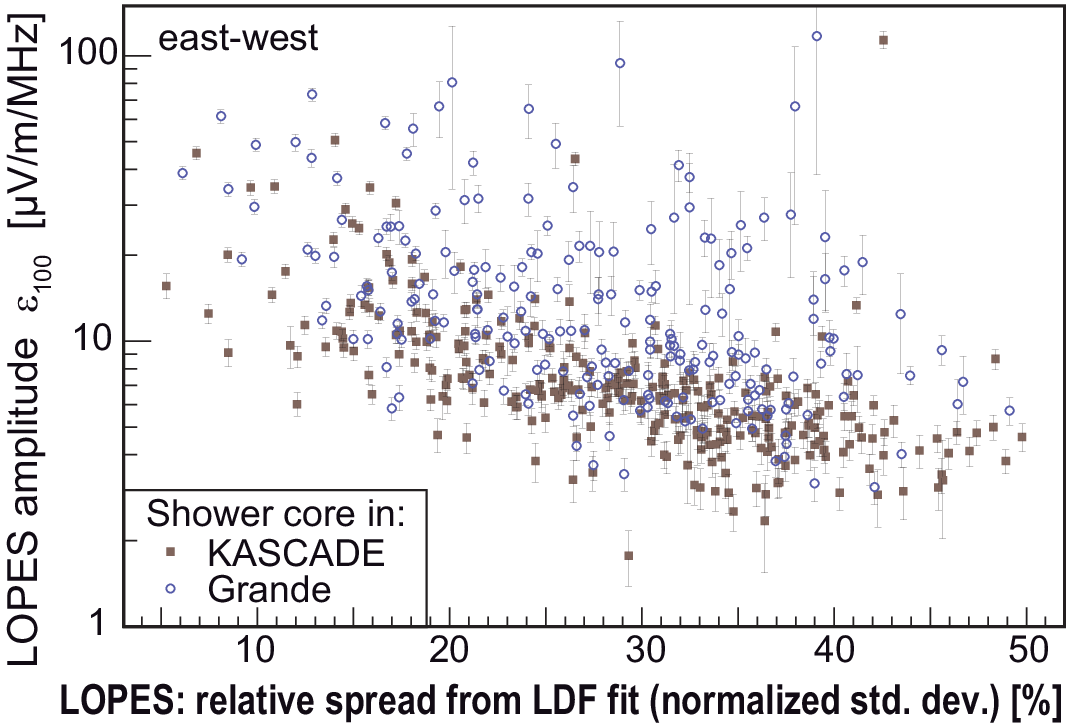}
\caption{Spread of the amplitudes in individual antennas around the exponential LDF fit for LOPES events with shower core in the KASCADE array and the Grande extension, respectively (cf. Fig.~\ref{fig_eventMap}). Most Grande events have larger uncertainties in $\epsilon_{100}$ than KASCADE events, since their shower core is at a larger distance to the LOPES antennas.}
\label{fig_dispersionFromLDFfitLOPES}
\end{figure} 

We checked whether a uniform exponential LDF is indeed a sufficient approximation for individual events, although the \lq real\rq~lateral distribution is more complex. Therefore, we studied the spread of the amplitudes measured in individual antennas around the exponential LDF fit. If the exponential LDF were a perfect description of the measurements, we would expect a spread only due to measurement uncertainties. In this case the spread is expected to decrease with rising signal-to-noise ratio, and thus with rising amplitude of the event. The spread indeed decreases with increasing amplitude (Fig.~\ref{fig_dispersionFromLDFfitLOPES}), however, it does not completely vanish since the exponential LDF is only a simplification. Still, for most events the spread seems to be dominated by noise, and the remaining spread at high amplitudes is small. Therefore, a uniform exponential LDF is justified for the purpose of comparing simulations to LOPES measurements, and the uncertainties introduced by this 
simplification will be small compared to other uncertainties.

\begin{figure}
\centering
\includegraphics[width=0.48\textwidth]{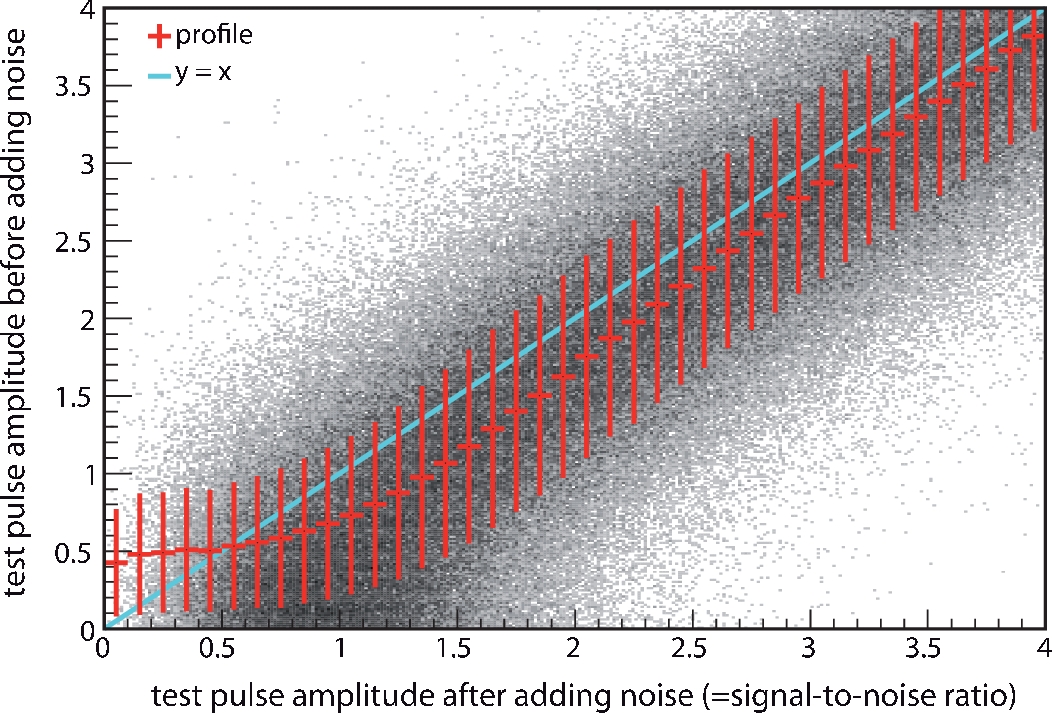}
\caption{True amplitude of test pulses (before adding noise) as function of the test pulse amplitude after adding noise to the pulses. All amplitudes are normalized to the noise level. Thus, the amplitude after adding noise is the signal-to-noise ratio. The profile reflects the mean change of the amplitude due to noise, and the standard deviation (vertical bars) is used as a measure for the uncertainty of an amplitude measurement at a certain signal-to-noise ratio.} \label{fig_noiseDependency}
\end{figure}

\subsection{Uncertainties}

Numerous systematic effects have been studied to check whether they influence the amplitude or slope parameter of the lateral distribution. The three main sources of uncertainties are the high ambient radio background at the site of the LOPES experiment, the calibration uncertainty of the absolute amplitude scale, and the energy uncertainty of the KASCADE-Grande reconstruction used as input for the REAS 3.11 and CoREAS simulations, respectively. All other studied effects can either be avoided or are negligible as summarized in the following subsections. However, there remain two uncertainties which are difficult to quantify: first, the simplification in the conversion from measured pulse amplitudes to electrical field strengths, since the used conversion formula is exact only for east-west polarized signals; second, the simplified treatment of the simulations by just applying a rectangular band filter to the effective LOPES bandwidth instead of a full detector simulation which is not available, yet. 
Nevertheless, we tested that our main conclusions do not depend on these two simplifications: first, all statements on the compatibility between the measurements and the simulations are also true when only events with arrival directions are selected for which the simplified treatment gives approximately the true east-west polarization component; second, with a preliminary version of a full detector simulation, we have tested that ignoring the frequency dependence of the experimental hardware properties will only have a slight systematic effect on the slope of the lateral distributions. This might be important when studying the composition of the primary cosmic rays, but it is not important when testing if simulations are compatible to measurements. A summary of all considered uncertainties can be found in table \ref{tab_summaryOfUncertainties}.

\begin{table*}
\centering
\caption{Summary of uncertainties (relative values) which are considered for the comparison of measured and simulated lateral distributions. Some effects are thought to have a negligible effect on the slope parameter $\eta$, which is indicated by '-'.} \label{tab_summaryOfUncertainties}
\vspace{0.1 cm}
\begin{tabular}{lcc} 
 & ~~~ uncertainty on $\epsilon_{100}$ ~~~  & ~~~ uncertainty on $\eta$ ~~~ \\
\hline
\textbf{Effects affecting the individual event}   & &\\ 
Noise                                                & \multicolumn{2}{c}{determined for each event by LDF fit}\\
Energy uncertainty per event                          & $20\,\%$& - \\
Amplitude calibration (environmental effects)~~~~~~~~& $5\,\%$ & - \\
\hline
\textbf{Effects affecting the absolute amplitude scale}              & &\\
Amplitude calibration                                & $35\,\%$ & - \\
Energy scale uncertainty (KASCADE-Grande)            & $20\,\%$ & - \\
Noise: average corrected systematic effect           & $4-5\,\%$ & $7-8\,\%$ \\
Pulse distortion due to dispersion                   & $< 5\,\%$ & $< 2\,\%$ \\
Up-Sampling                                          & $< 2\,\%$ & $< 2\,\%$ \\
\hline
\end{tabular}
\end{table*}

\subsubsection{Noise}
In the noisy environment of a large research center understanding the influence of noise on the measurements is important to avoid any biases on the reconstruction of lateral distributions \cite{SchroederARENA2010}. Noise generates an uncertainty for amplitude measurements which is reflected in the statistical uncertainties of the lateral distribution fit (minimizing $\chi^2$-fit done with ROOT 5.30.06 \cite{ROOT}). Understanding the influence of noise on radio measurements in more detail requires some effort: Radio noise can in principle interfere constructively or destructively with the radio signal and thus increase or decrease the true signal. This effect has been studied by adding measured noise in the analysis software to measured test pulses. Depending on the signal-to-noise ratio, an increase or decrease is not equally likely (see Fig.~\ref{fig_noiseDependency}). For details on the test pulse study and on the method how the noise level is determined in a consistent way see Ref.~\cite{SchroederPhD}. 
Correcting for the noise influence is important, since noise systematically flattens lateral distributions, because antennas at larger distances typically have a lower signal-to-noise ratio than antennas close to the shower axis. The resulting effect on $\eta$ is an average decrease due to noise by about $7-8\,\%$, and the effect on $\epsilon_{100}$ is an average increase by about $4-5\,\%$. To avoid these biases, every single amplitude measurement in each individual antenna is corrected for the mean effect of noise observed at the corresponding signal-to-noise ratio (profile in Fig.~\ref{fig_noiseDependency}).

\subsubsection{Amplitude calibration}
The uncertainties of the absolute amplitude calibration are discussed in detail in Ref.~\cite{NehlsHakenjosArts2007}. However, all uncertainties in Ref.~\cite{NehlsHakenjosArts2007} are given for the gain factor which is proportional to the square of the amplitude. The systematic calibration uncertainty originates from different sources, which contribute to this analysis in different ways. The largest uncertainty is the gain uncertainty of the calibration antenna. It is given by the manufacturer as $2.5\,$dB which corresponds to a scale uncertainty of about $35\,\%$ for the radio amplitude. The contribution of the LOPES antenna model to the scale uncertainty is significantly smaller than $35\,\%$ and thus neglected for this analysis. Although a scale uncertainty is not important when comparing several LOPES events with each other (as in Ref.~\cite{ApelArteagaAsch2010}), it is crucial when comparing LOPES events with other experiments or simulations. In particular, the scale uncertainty limits the testing 
power whether a theoretical model can predict the absolute amplitude correctly. Even for a perfect simulation, the measured and simulated amplitude (i.e.~$\epsilon_{100}$) could deviate by up to $35\,\%$, though by the same amount and in the same direction for all events.

In addition, there are two other types of calibration uncertainties, mainly due to the unknown influence of environmental variations. First, the gain of all antennas can change simultaneously in the same direction, since all antennas are subject to more-or-less the same environmental conditions at the same time. This results in a per-event uncertainty of $\epsilon_{100}$ which can be treated in a similar way as a statistical uncertainty and thus is quadratically added to the statistical uncertainty of the LDF fit. Second, since not all antennas react exactly equally to environmental changes, there is an additional uncertainty for the amplitude measurement in each individual antenna which has to be quadratically added to the uncertainty due to noise. Both types of uncertainties are almost an order of magnitude smaller than the scale uncertainty. However, an underestimation of either of the two types of uncertainties might lead to a wrong rejection of the theoretical model. Thus, we conservatively estimate 
both type of amplitude uncertainties to $5\,\%$ which is slightly more than the total environmental uncertainty stated in Ref.~\cite{NehlsHakenjosArts2007} ($9\,\%$ for the gain corresponding to $4.5\,\%$ for the amplitude).

\subsubsection{Energy uncertainty}
For the KASCADE-Grande energy reconstruction we use formulas based on the measurement of the muon and electron numbers of the KASCADE array \cite{Glasstetter2006} and the numbers of muons and charged particles for KASCADE-Grande \cite{Wommer2007, ApelGrandeEnergySpectrum2012}. The uncertainties of this energy reconstruction formulas have been studied with Monte Carlo simulations. For the events of the LOPES selection, the typical uncertainty is about $20\,\%$ for individual events. Since the radio amplitude scales almost linearly with the energy, this means a corresponding systematic uncertainty of $\epsilon_{100}$ for the simulations, which we add quadratically to the statistical uncertainty when comparing $\epsilon_{100}$ of LOPES and the simulations event-by-event. In addition, the energy reconstruction of KASCADE, respectively, KASCADE-Grande has a scale uncertainty estimated to about $20\,\%$. The effect of this scale uncertainty is similar to the amplitude calibration scale uncertainty, i.e., all 
simulated amplitudes could be simultaneously by up-to $20\,\%$ too high or too low. For the slope parameter $\eta$, the energy uncertainty is neglected. First, the energy dependence of $\eta$ is expected to be very weak and, second, only the mean $\eta$ is compared between measurements and simulations, since a comparison of $\eta$ for individual events is hampered by shower-to-shower fluctuations.

\subsubsection{Antenna altitude and shower inclination}
All LOPES antennas are placed within $1\,$m at the same altitude. Therefore, any possible effects of the altitude on the measured lateral distributions can be neglected to very good approximation. However, for inclined showers the antennas have a different altitude in shower coordinates, i.e.~relative to a plane perpendicular to the shower axis. Thus, the air shower radio signal can have different propagation lengths to antennas at an equal distance $d$ to the shower axis. The order-of-magnitude of this effect can roughly be estimated by assuming a simple point source in the sky with a spherical wavefront. Although, the real radio wavefront of air showers is approximately conical \cite{SchroederICRC2011}, already the simple spherical approximation reveals that the effect of the shower inclination is negligible for LOPES. When assuming a distance of $5\,$km (typical radius when fitting a spherical wavefront during the beamforming analysis) and a zenith angle $\theta = 30^\circ$, the different antenna heights 
in shower coordinates lead to a maximum difference in the radio amplitude of $(200\,\mathrm{m} \cdot \sin{30^\circ}) / 5\,\mathrm{km} \approx 2\,\%$, since the lateral extension of LOPES is about $200\,$m. The typical effect ought to be even smaller. Thus, the systematic uncertainties due to the shower inclination can be neglected for LOPES, but might be important for radio arrays with larger lateral extensions.

\subsubsection{Pulse distortion (filter dispersion)}
Pulse distortion by the dispersion of the LOPES bandpass filter can increase or decrease the pulse amplitude depending on the pulse shape \cite{SchroederAschBaehren2010}. However, the pulse shape of measured air shower radio pulses is known only roughly, since the high ambient noise level at LOPES and the limited effective bandwidth ($43-74\,$MHz) do not allow precise pulse shape measurements. Thus, we try to decrease systematic uncertainties by digitally correcting the measured radio pulses for the known pulse distortion of the LOPES bandpass filters. All other LOPES components (e.g., the antennas) contribute less to pulse distortion, and we can take the correctable effect of the bandpass filter dispersion as an upper limit for the remaining pulse distortion of all other LOPES components. It has been studied by reconstructing the lateral distribution with and without digitally correcting for the dispersion of the bandpass filter: The remaining uncertainty of $\epsilon_{100}$ is smaller than $5\,\%$ and the 
uncertainty of $\eta$ is smaller than $2\,\%$.

\subsubsection{Up-Sampling}
Since the LOPES hardware fulfills the Nyquist sampling theorem by sampling in the second Nyquist domain ($80\,$MHz sampling rate and an effective bandwidth of $43-74\,$MHz), up-sampling is applied. Up-sampling is the correct interpolation between the sampled data points of the time series. To test the effect of up-sampling on the reconstruction of the lateral distributions, the same events have been analyzed with different up-sampling rates: up-sampling is mostly important for the cross-correlation beamforming which precedes the reconstruction of the lateral distribution. The mean effect on the lateral distribution itself is smaller than $2\,\%$ for both $\eta$ and $\epsilon_{100}$. In principle, the effect can be made even smaller by using a higher up-sampling factor. However, this would demand additional computing recourses, and is not necessary, since the total uncertainty on $\eta$ and $\epsilon_{100}$ is dominated by other effects.

\begin{figure*}[t]
 \centering
 \includegraphics[width=0.92\columnwidth]{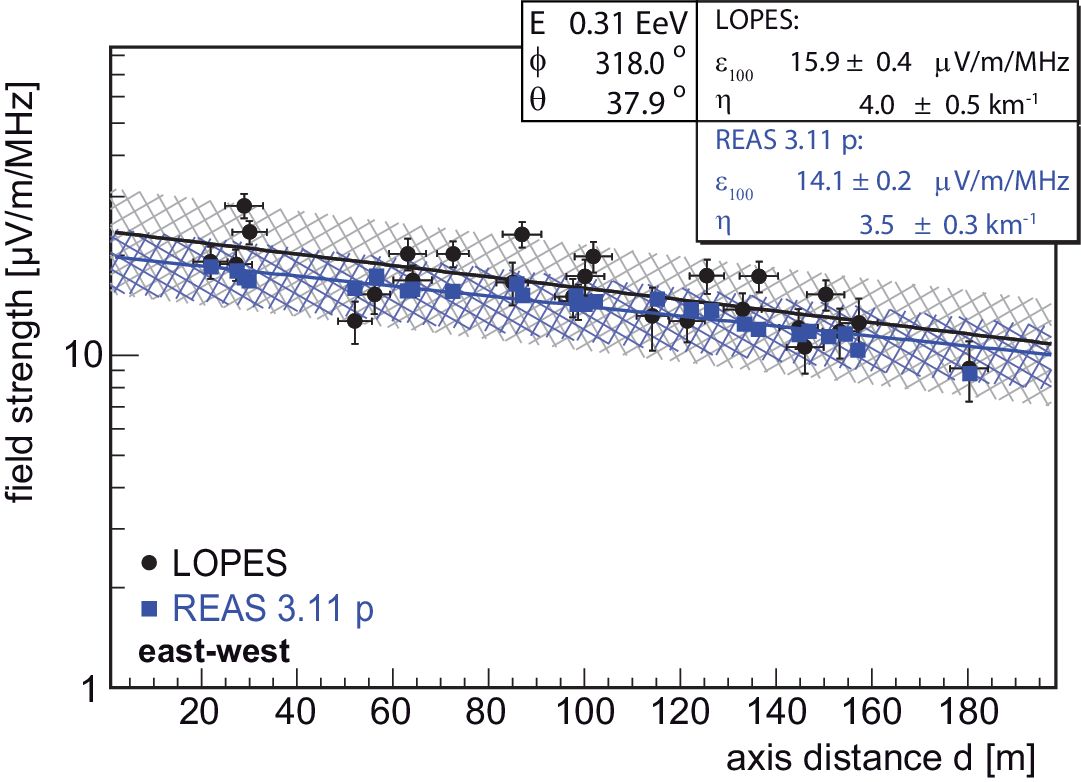}
  \hskip 0.13\columnwidth
 \includegraphics[width=0.92\columnwidth]{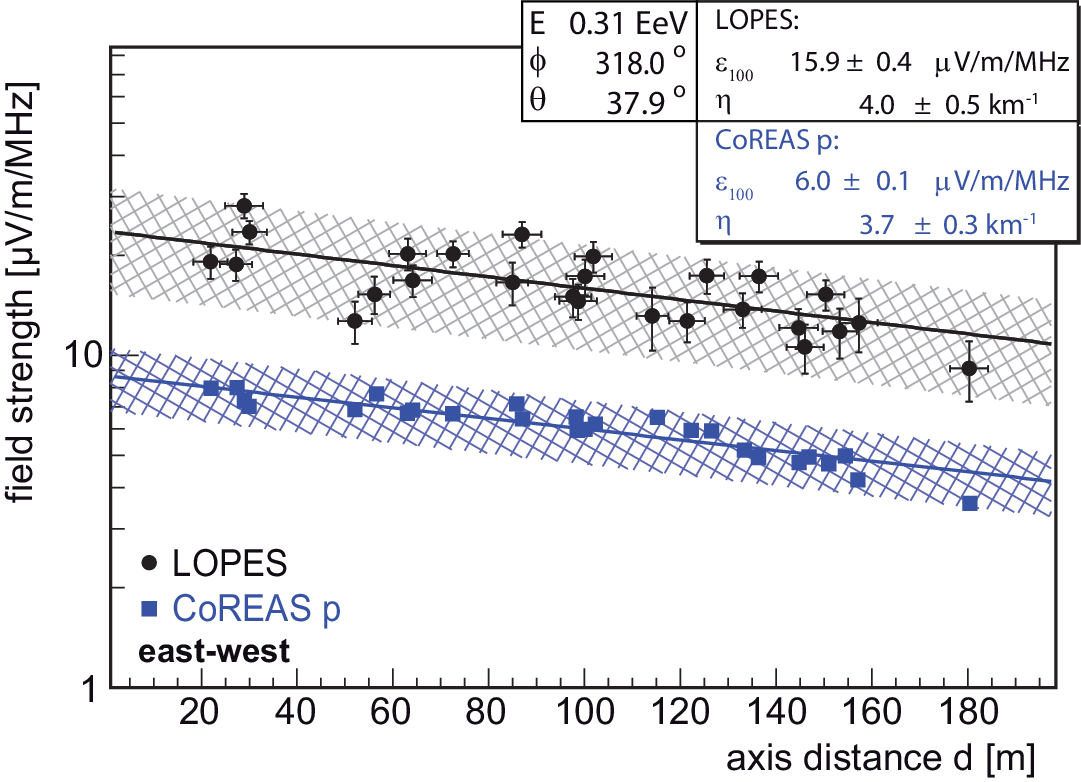}
  \vskip 0.2 cm
 \includegraphics[width=0.92\columnwidth]{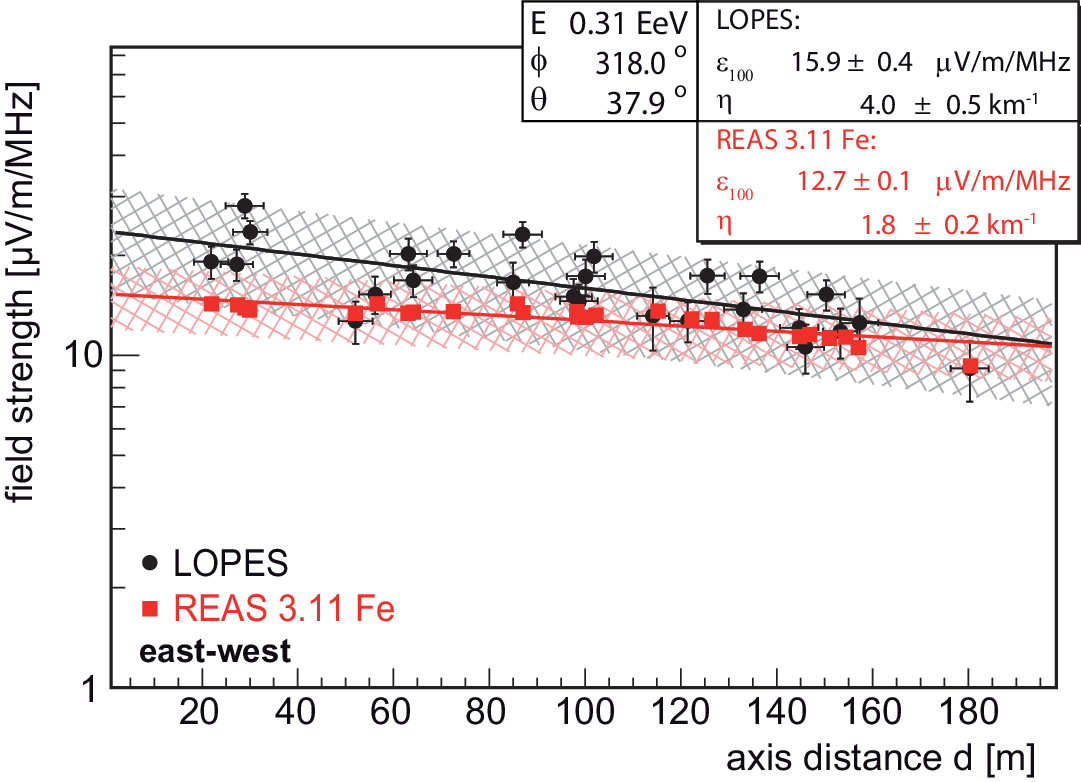}
  \hskip 0.13\columnwidth
 \includegraphics[width=0.92\columnwidth]{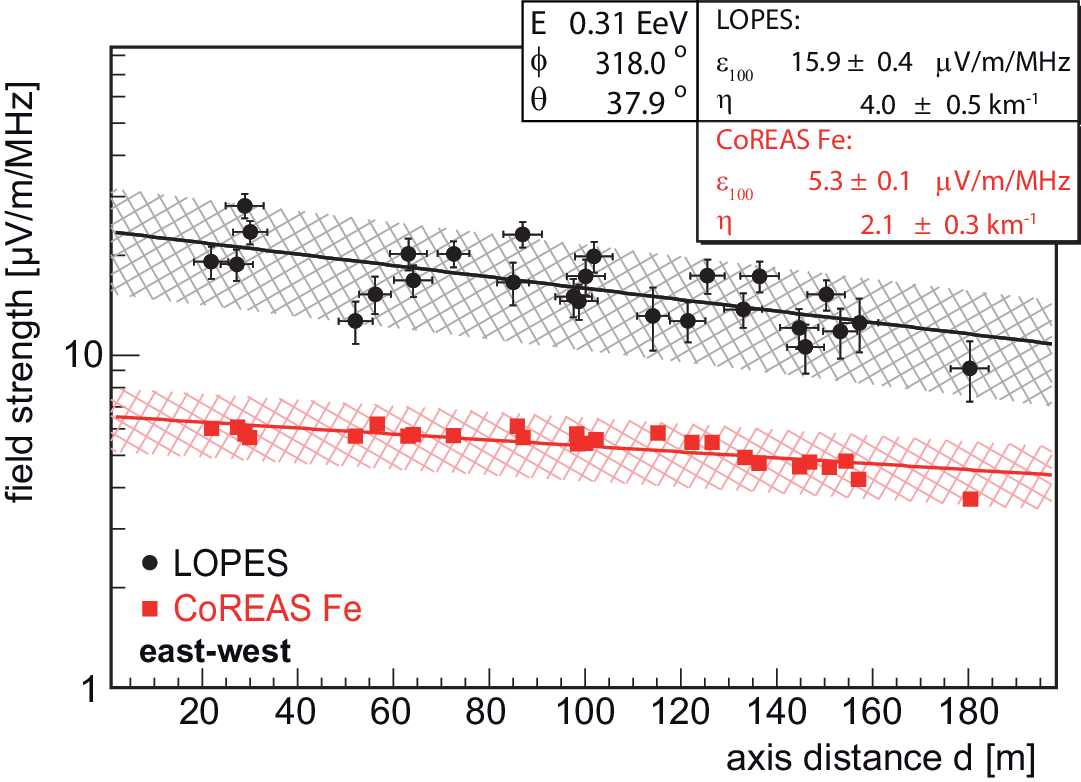}
  \caption{Example event for the comparison of LOPES lateral distributions to REAS 3.11 simulations (left), and CoREAS simulations (right), for protons (top) and iron nuclei (bottom) as primary particles: amplitudes at the individual antennas and an exponential fit. The stated uncertainties of $\epsilon_{100}$ and $\eta$ are the statistical fit uncertainties. The systematic uncertainties are given as bands: a $35\,\%$ scale uncertainty for the LOPES measurements and a systematic uncertainty of typically $20\,\%$ for each simulated event due to the energy uncertainty of KASCADE-Grande. For reference, also the energy $E$, the zenith angle $\theta$ and the azimuth angle $\phi$ are given.}
 \label{fig_REAS3exampleLateralDistributions}
\end{figure*}

\section{Simulations}
\label{sec_simulations}
For the comparison of the lateral distribution of LOPES measurements with the lateral distribution predicted by simulations, REAS (v 3.11) and CoREAS (v 1.0) were selected to model the radio emission from air showers. Both are publicly available Monte Carlo simulation codes developed in the frame of LOPES. In principle, both simulate the same physics, i.e., they calculate the radio emission from air showers by superposing the radiation of the individual particles of the air shower. Both, the established REAS simulations and the new CoREAS simulations, are based on the endpoint formalism, i.e. no specific assumptions on the emission mechanism have to be made \cite{LudwigHuege2010, James2011}.

In principle, any radiation originating from the acceleration, creation, annihilation or disappearance of air-shower particles can be described. In practice, the air showers are simulated with CORSIKA \cite{HeckKnappCapdevielle1998} using the hadronic interaction models QGSJetII.03 \cite{OstapchenkoQGSjetII2006} and FLUKA \cite{BattistoniFLUKA2007, FerrariFLUKA2005}. Other hadronic interaction models might result in a slightly different radio emission, depending on how much the electromagnetic air-shower component changes. Since only the relativistic air-shower electrons and positrons have a significant effect on the simulated radio emission \cite{LudwigPhD}, other particles are neglected. Moreover, acceleration by electric fields is neglected, since experimental results indicate that atmospheric electric fields do not have any significant effect on the radio emission during normal weather conditions \cite{BuitinkApelAsch2006, LopesThunderstormARS2011}. Finally, REAS 3.11 and CoREAS both consider a realistic,
 height-dependent refractive index of the atmosphere, which changes the coherence conditions for any radio emission generated by the air shower.

The difference between REAS 3.11 and CoREAS is that CoREAS is directly implemented in CORSIKA, while REAS 3.11 is a separate software which makes use of output generated by CORSIKA. For REAS, the information on the electrons and positrons is taken from histograms, thus losing information on the individual particles. Consequently, CoREAS is expected to provide a more accurate simulation of the same physics. Nevertheless, we see in the performed comparisons that REAS 3.11 gives a better description of the LOPES data, and thus compare to both simulation codes.

For the comparison between the simulations and LOPES data as shown in this article, each event was simulated for two different types of the primary particle of the air shower, namely protons and iron nuclei (see Fig.~\ref{fig_REAS3exampleLateralDistributions} for example lateral distributions). For this, a simulated air shower reproducing the observables measured by KASCADE is chosen, in particular the geometry, the energy and the muon number. Furthermore, before comparing the output of the simulations with the data obtained with LOPES, a frequency filter needs to be applied to the simulations since they are performed for an unlimited frequency band and LOPES is measuring in the effective band of $43-74\,$MHz. The filtering of both the REAS and the CoREAS simulations to a finite observing bandwidth is performed with REASPlot (a helper application included in the REAS package) using an idealized rectangle filter.

\begin{figure*}[t]
\centering
\includegraphics[width=0.65\textwidth]{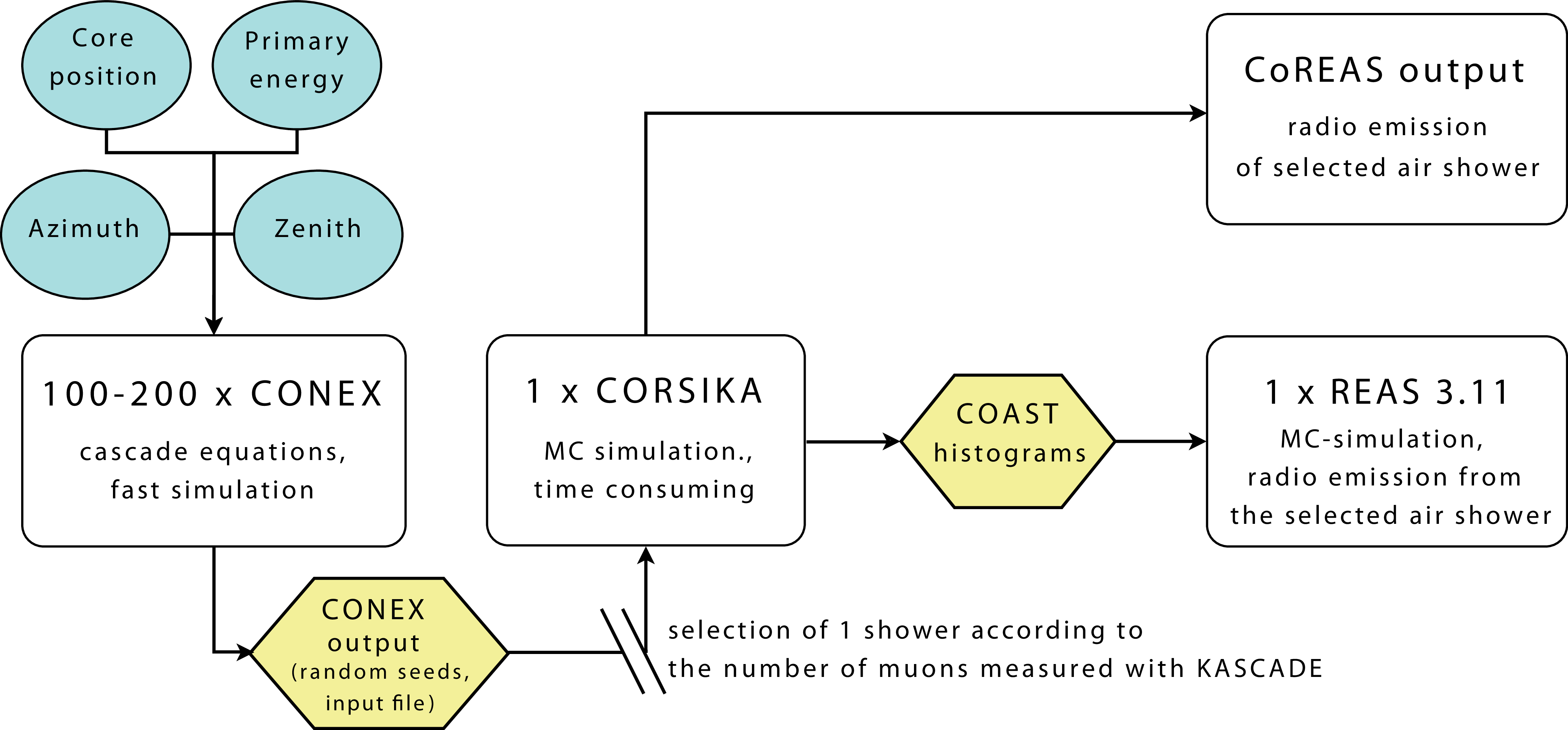}
\caption{Flow chart of the simulation chain for the production of the REAS 3.11 and CoREAS simulations used in this article. In the ellipses are the input parameters determining the air shower physics. In the hexagons are parameters or information from the programs themselves for the following simulation program.} \label{fig_simchain_muon}
\end{figure*}

\subsection{Selection of CORSIKA shower by muon number}
\label{sec_simSelection}
For the purpose of testing simulation codes against data, it would be ideal if one could reproduce the experimentally observed shower with all its properties. For the geometry and the energy this is relatively easy, since these properties can be selected as input parameters for the shower simulations, in our case CORSIKA. However, due to shower-to-shower fluctuations, each simulated shower (with the same energy and geometry) is different. This leads to systematic uncertainties when comparing simulations and measurements event-by-event, and this is the reason why we give quantitative conclusions based on histograms and average values.

To select one of the numerous air-shower simulations which are possible for each event, we developed a new methodology taking into account the measurements of the KASCADE particle detector array, i.e., in addition to the shower geometry and energy, in particular the muon number. A selection of a specific shower for the radio simulation is necessary, since only a few air-shower and REAS 3.11 simulations per measured cosmic-ray event are processable due to limited computing times. Of the many possible simulated showers for the geometry and energy given by KASCADE-Grande, we decided to choose a shower with a muon number similar to the measurement, instead of a random shower.

The methodology we developed for the shower selection is summarized in Fig.~\ref{fig_simchain_muon}. It makes use of CONEX \cite{BergmannCONEX2007, PierogICRC2011}, an air-shower simulation code incorporated in the used version of CORSIKA. CONEX saves computing time compared to a standard CORSIKA simulation, since it uses a Monte Carlo simulation only for the first few air-shower interactions, and then makes use of cascade-equations to calculate the air shower. Thus, with CONEX is is possible to simulate several 100 showers within reasonable computing time, and then select one of these showers for the complete CORSIKA simulation.

The method for shower selection might be especially useful for multi-hybrid observatories, it can also be used to select showers by $X_\textrm{max}$ instead of by muon number. Therefore, we give a detailed description of the method in Appendix B. It can be used to improve the method and transfer it to other experiments.

If a measurement of additional, complementary shower parameters were available (e.g., $X_\textrm{max}$ and muon number), then in principle it should be possible to beat shower-to-shower fluctuations by selecting a simulated shower which reproduces all measured shower parameters. It is clear, that in the present analysis we can only partially reach this goal, because KASCADE-Grande features only a measurement of the muon number, but no measurement of $X_\textrm{max}$. In addition, there are systematic uncertainties, since there are no hadronic models available which completely reproduce all measured properties of air-shower muons. For these reasons, the shower selection by muon number has on average only a small effect on the impact of shower-to-shower fluctuations, at least in the way we performed it (see Appendix B for details). In particular, the average values and the standard deviations of the slope parameter $\eta$ and the amplitude parameter $\epsilon_{100}$ change only marginally when the showers are 
selected by muon number instead of randomly. Still, for principle reasons we want to use all observables provided by KASCADE-Grande. Consequently, we decided to use the muon selection, although it has no significant effect on the conclusions of the analysis. In any case, such a selection criterion might be useful for future studies at other experiments.

\begin{figure*}[p!]
  \centering
  \includegraphics[width=0.77\columnwidth]{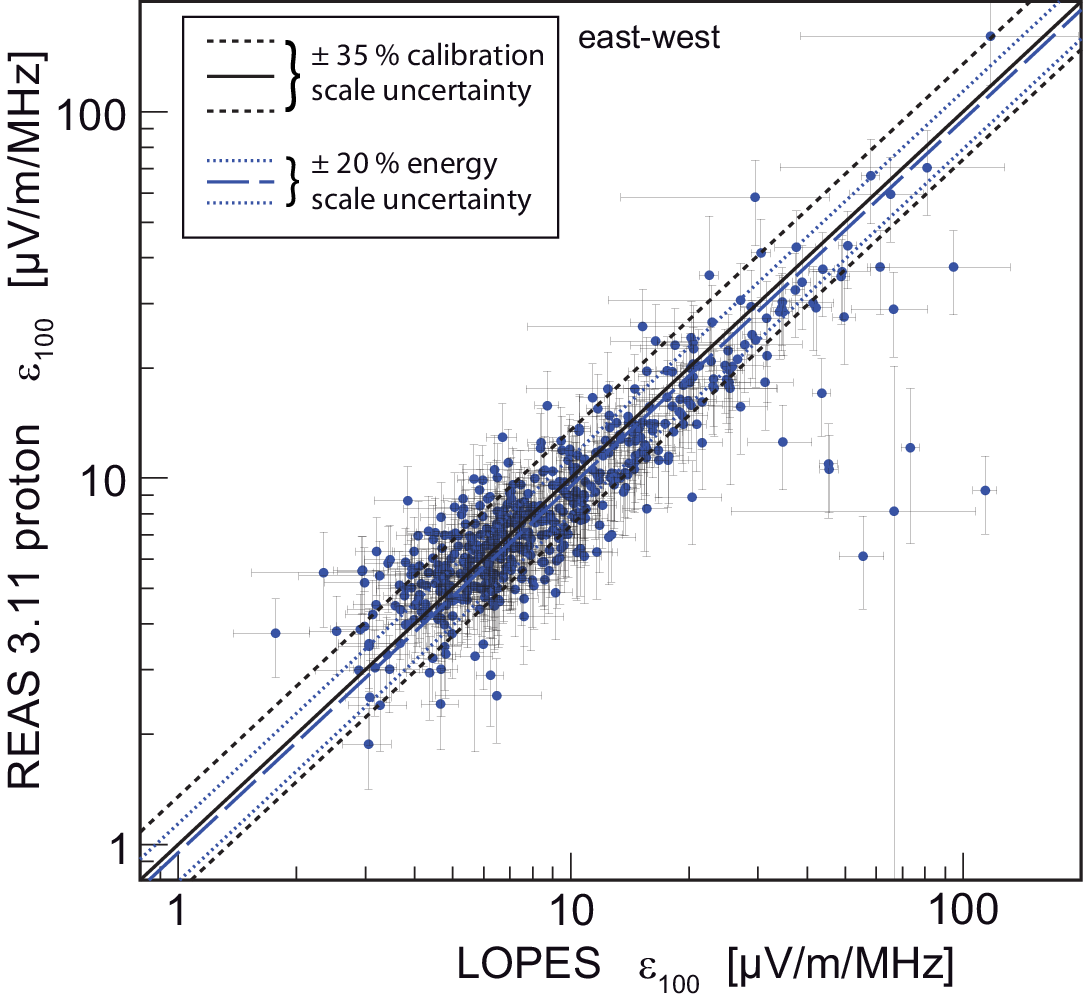}
  \hskip 0.25\columnwidth
  \includegraphics[width=0.77\columnwidth]{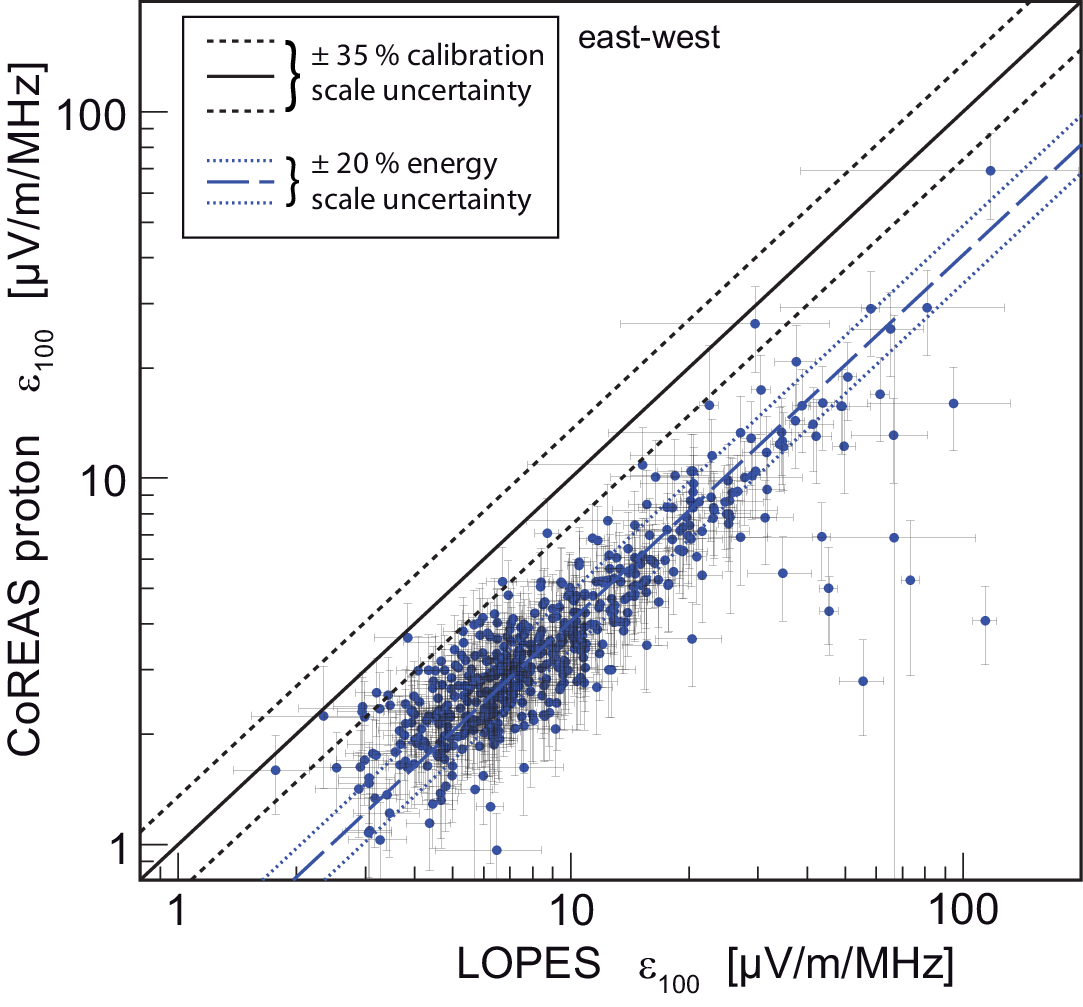}

  \includegraphics[width=0.77\columnwidth]{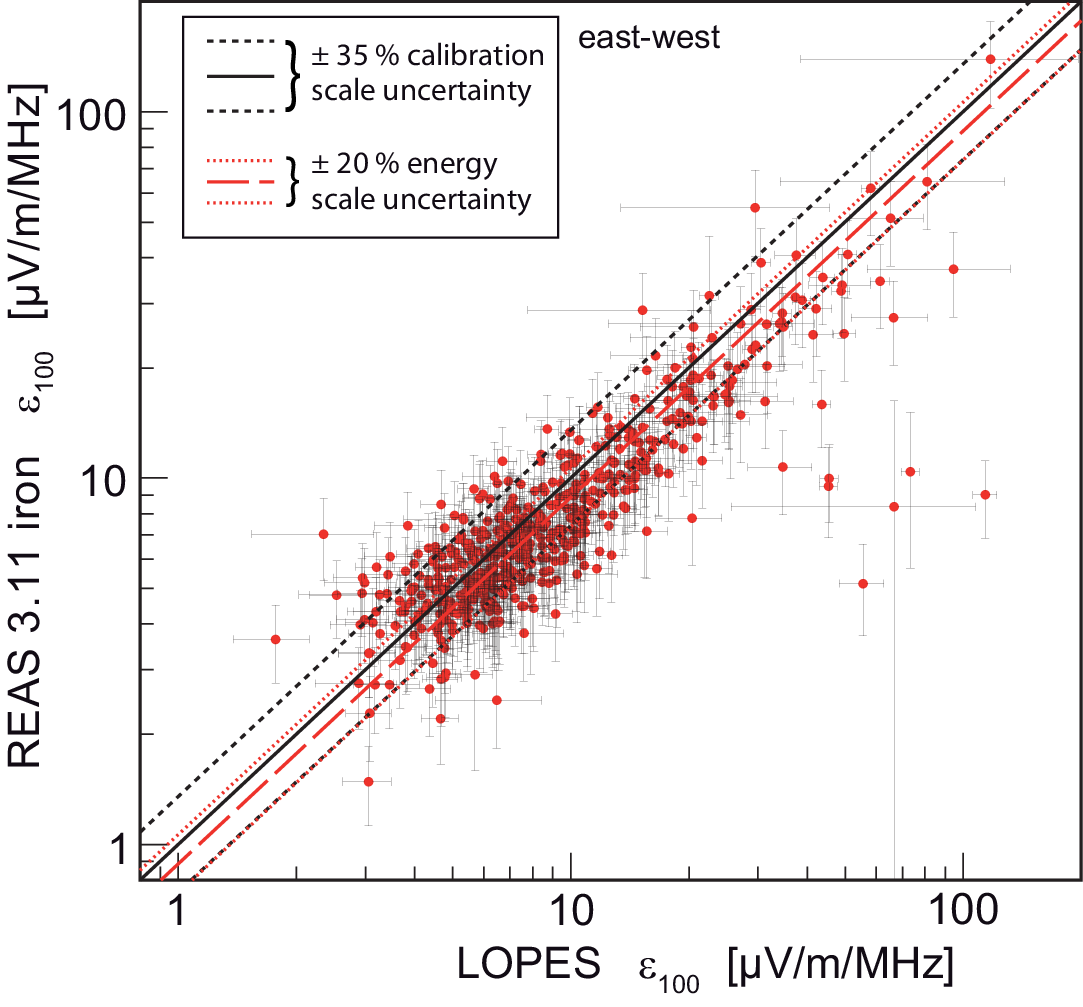}
  \hskip 0.25\columnwidth
  \includegraphics[width=0.77\columnwidth]{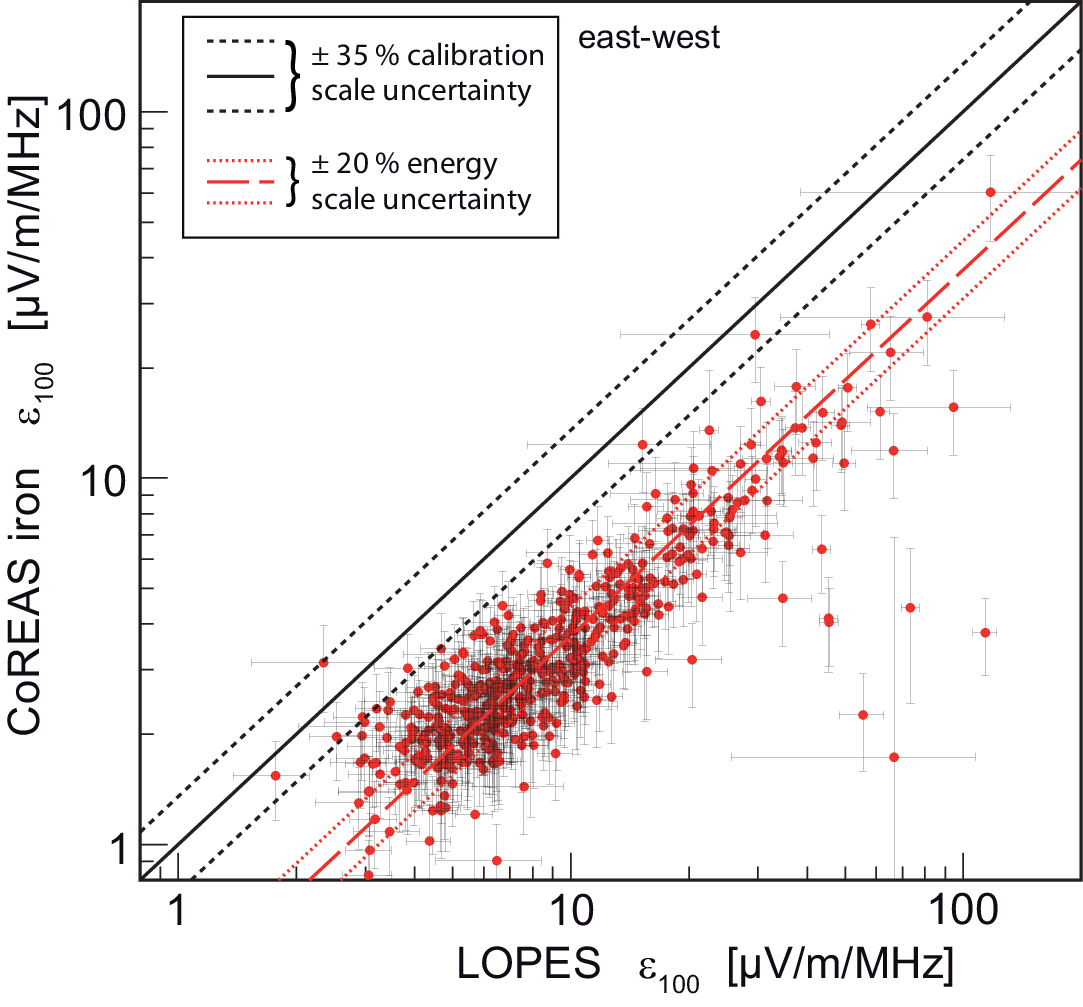}
  \caption{Per event comparison of the amplitude parameter $\epsilon_{100}$ between LOPES measurements and REAS 3.11 (left) and CoREAS (right), respectively, simulations for protons (top) and iron nuclei (bottom) as primary particles. The dashed lines around the solid line indicate the scale uncertainty of $\pm 35\,\%$ due to the absolute amplitude calibration of LOPES. The dotted lines around the long-dashed line indicate the $\pm 20$\,\% scale uncertainty of the KASCADE-Grande energy reconstruction. These lines have been shifted by the scale mismatch between the simulations and the measurements (see Fig.~\ref{fig_spread} for the scale factors).}
   \label{fig_amplitudeComparison}
\end{figure*}

\begin{figure*}[p!]
\centering
  \includegraphics[width=0.92\columnwidth]{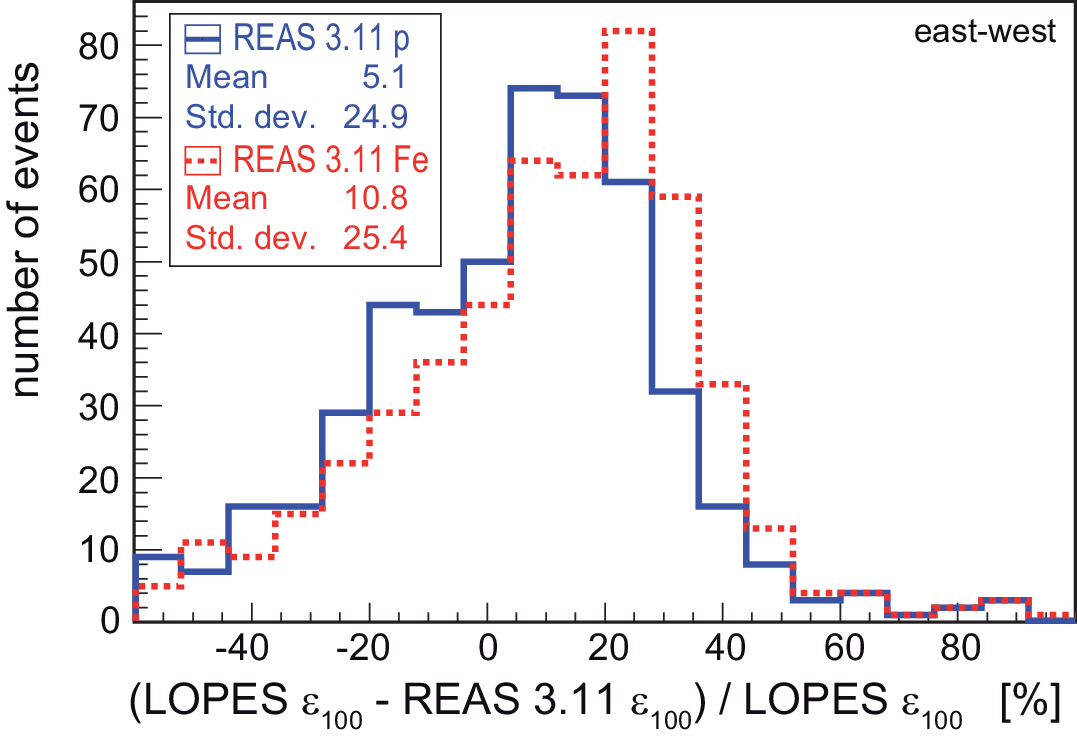}
  \hskip 0.12\columnwidth
  \includegraphics[width=0.92\columnwidth]{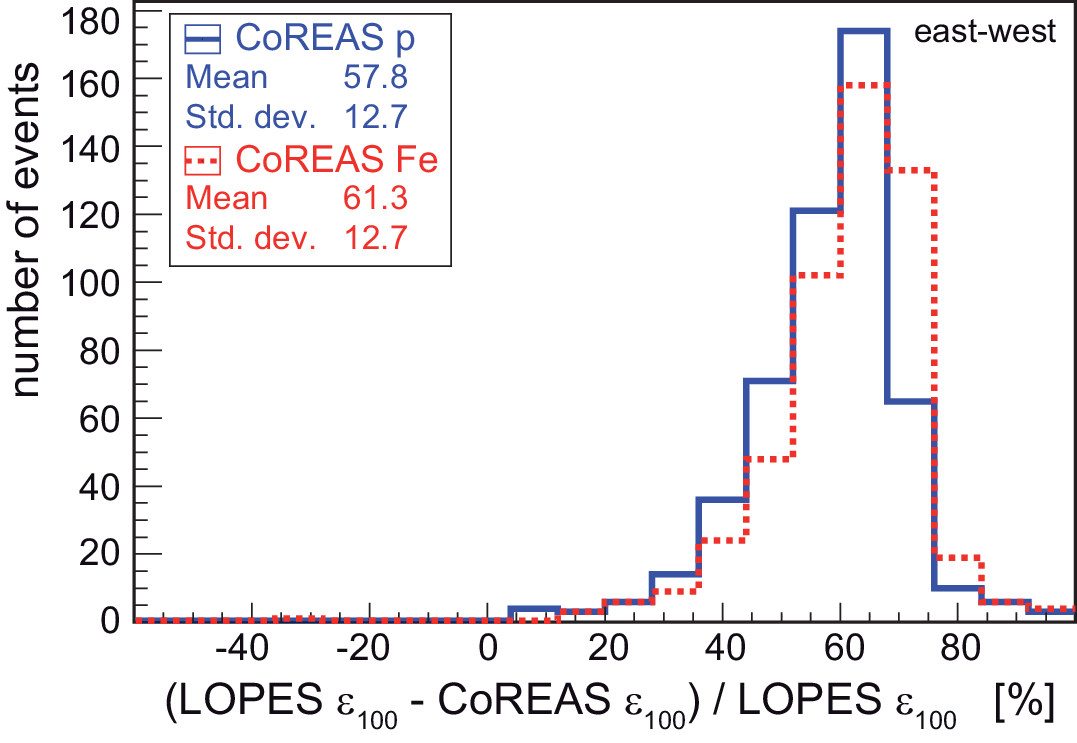}
\caption{Relative deviation of $\epsilon_{100}$. For REAS 3.11 the mean deviation is significantly smaller than the calibration scale uncertainty of $35\,\%$, which, however, is not the case for CoREAS.} \label{fig_histEpsDev}
\end{figure*}

With this, the computing time is optimized while taking into account all shower information available from the KASCADE measurements and selecting a suitable air shower and its radio emission for the comparison. The selection criterion based on the number of muons measured by KASCADE has the advantage that the specific air shower represents the measured air shower with a higher probability than a typical or randomly chosen air shower. However, even when fixing the muon number, there are still shower-to-shower fluctuations. To completely avoid the effect of shower-to-shower fluctuations, one would have to fix at least another shower parameter, e.g., the atmospheric depth of the shower maximum, $X_\textrm{max}$. This is not possible for LOPES, since KASCADE features no $X_\textrm{max}$ measurement. Consequently, the simulation of a single event is not expected to exactly reproduce the measurement of that individual event. Nevertheless, the simulations are expected to reproduce the measurements on average and 
the statistics of simulated and measured events are sufficiently large to test this.

\begin{figure*}
\centering
\includegraphics[width=0.92\columnwidth]{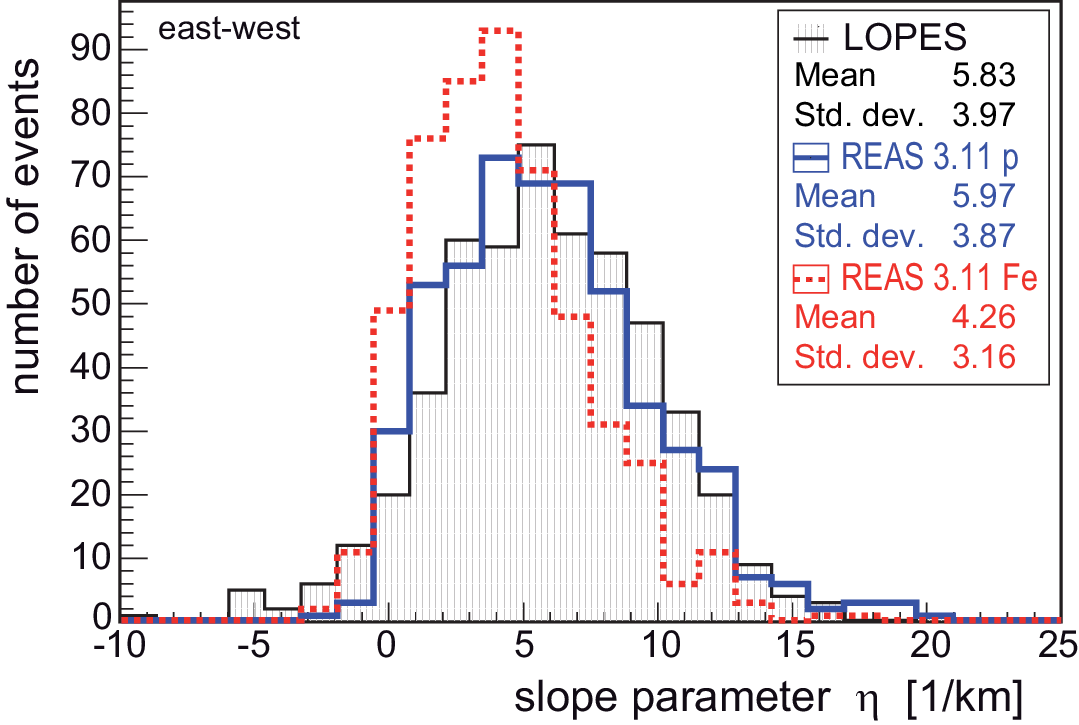}
\hskip 0.12\columnwidth
\includegraphics[width=0.92\columnwidth]{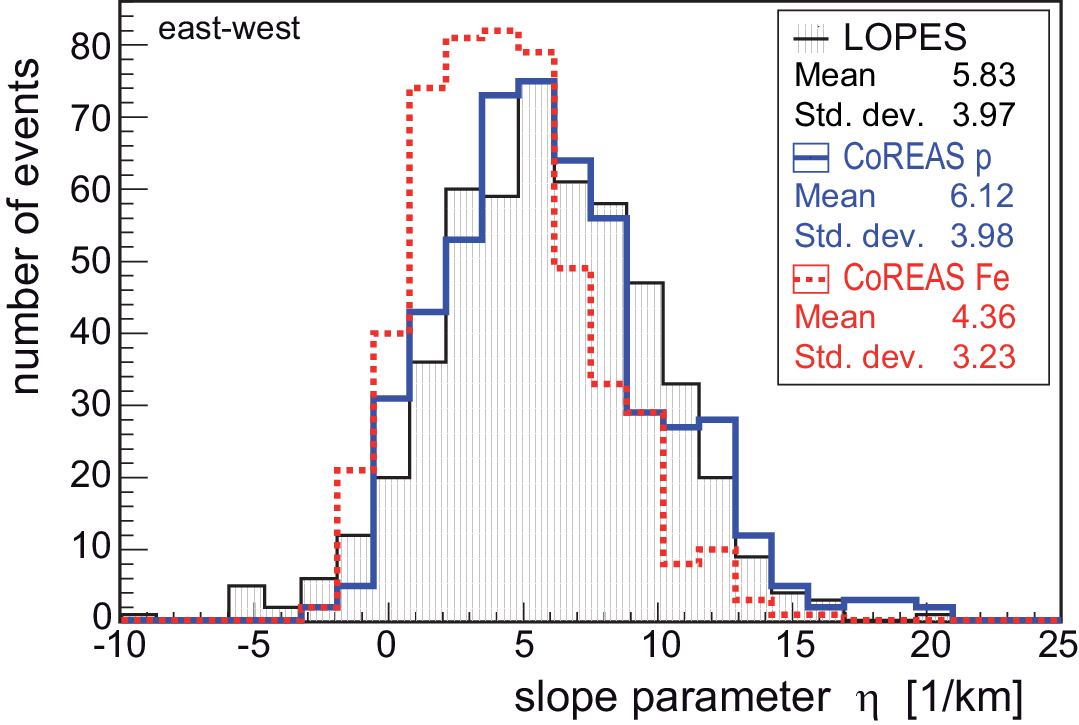}
\caption{Comparison of the slope parameter $\eta$ between LOPES measurements and REAS 3.11, respectively, CoREAS simulations for protons and iron nuclei as primary particles.} \label{fig_slopeComparison}
\end{figure*}

\begin{figure*}
\centering
\includegraphics[width=0.92\columnwidth]{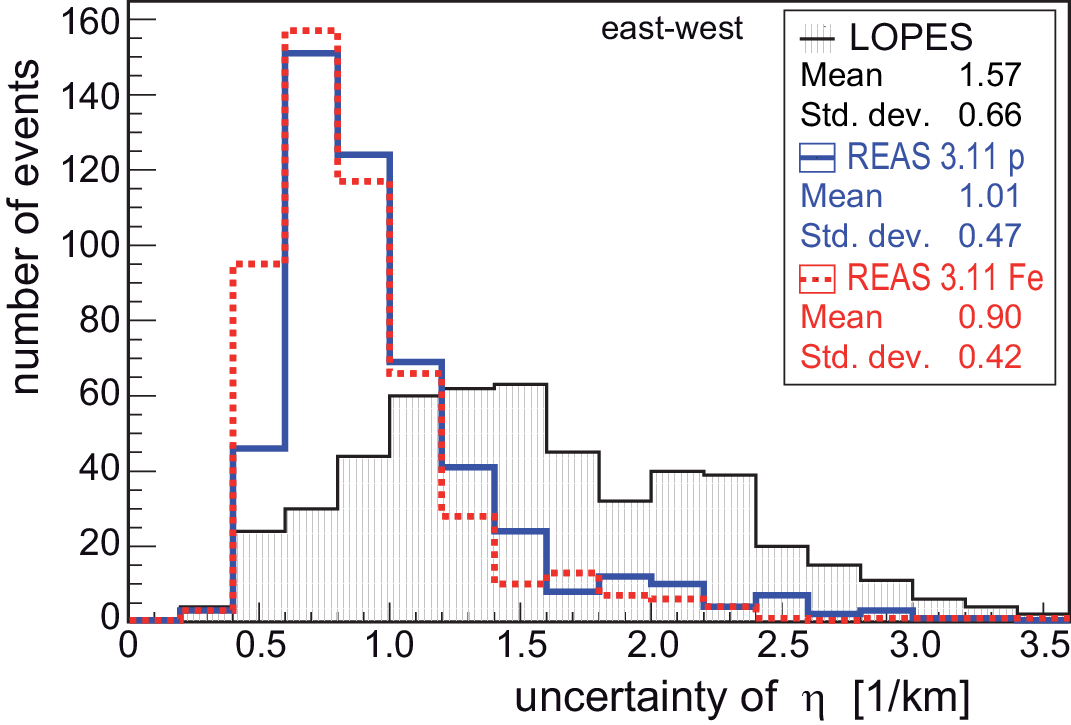}
\hskip 0.12\columnwidth
\includegraphics[width=0.92\columnwidth]{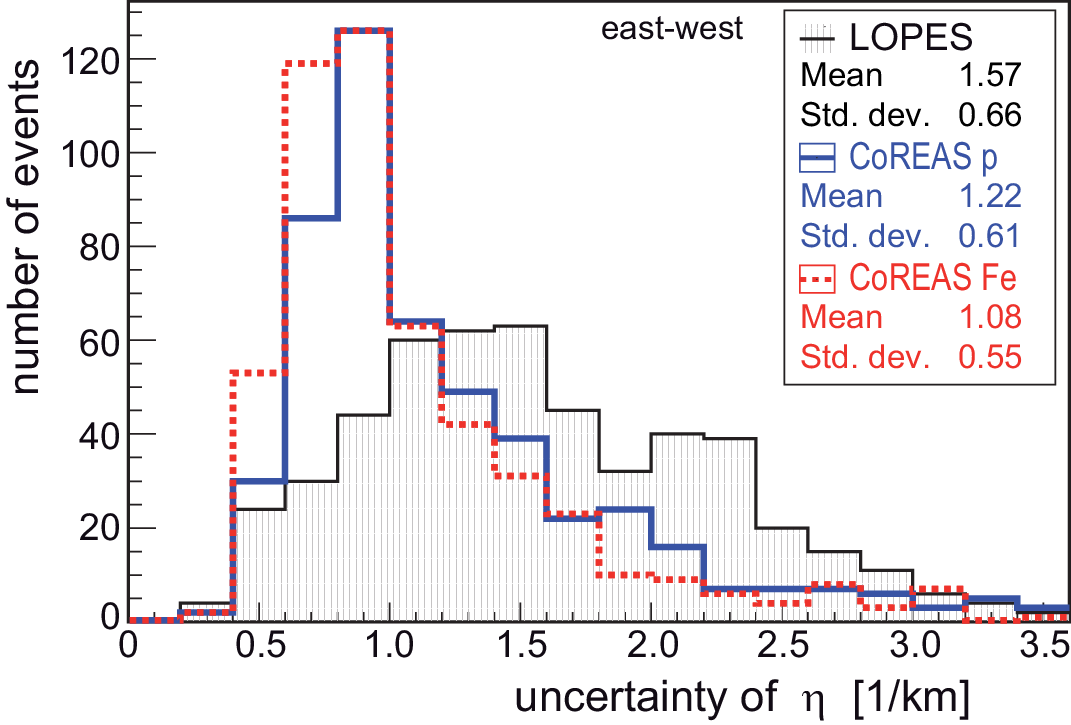}
\caption{Statistical uncertainty of the slope parameter $\eta$. The uncertainties of the LOPES measurements are generally larger than for the simulations, since the measurements are affected by noise, but the simulations do only contain numerical noise.} \label{fig_slopeUncertainty}
\end{figure*}

\begin{figure*}
\centering
\includegraphics[width=0.92\columnwidth]{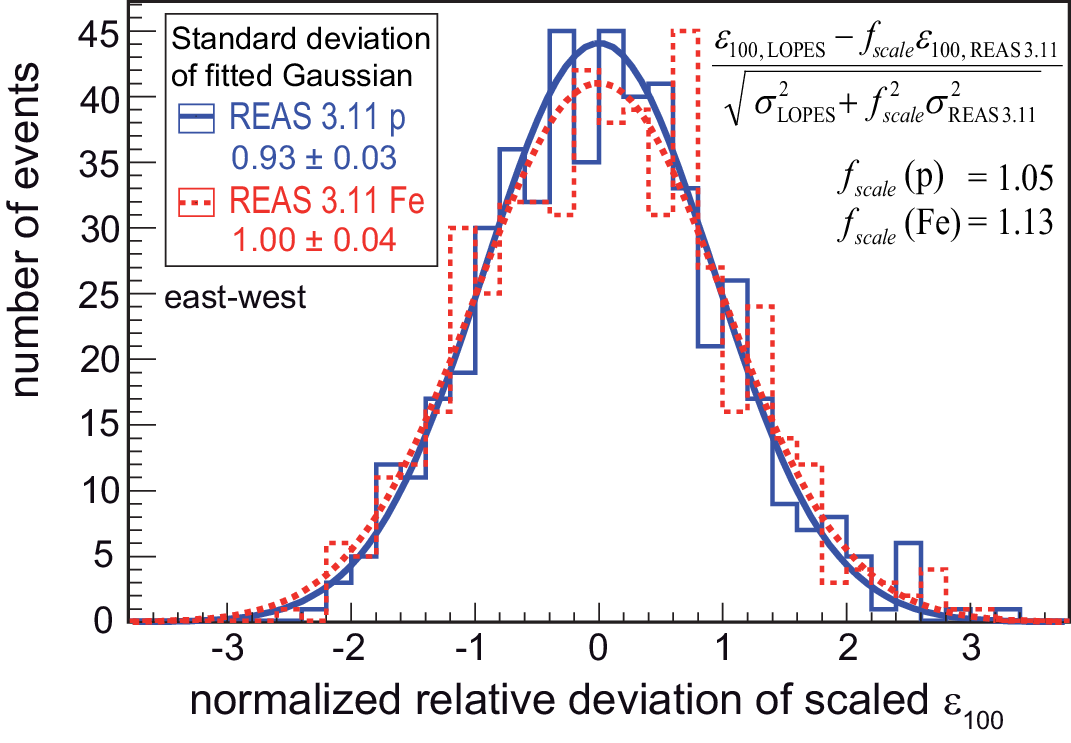}
\hskip 0.12\columnwidth
\includegraphics[width=0.92\columnwidth]{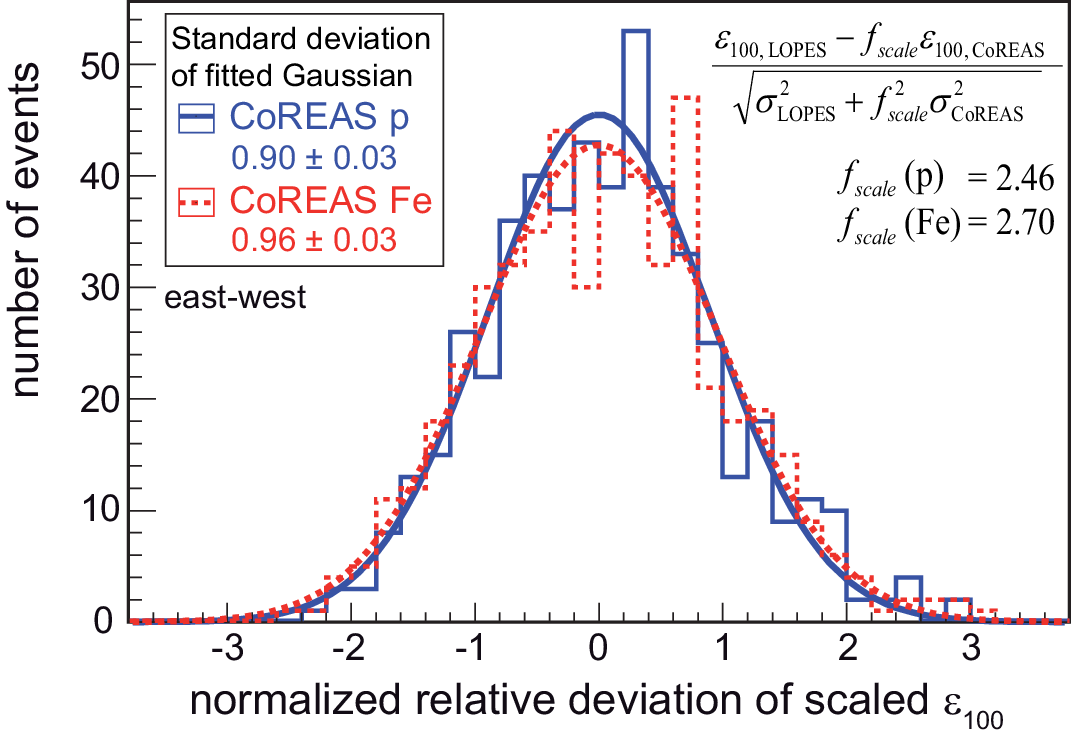}
\caption{Deviation between the measured $\epsilon_{100}$ and the scaled simulated $\epsilon_{100}$ divided by the total uncertainty of the individual events, and a fitted Gaussian for each histogram. The scaling factor $f_\mathrm{scale}$ has been chosen such that the fitted Gaussian is centered around 0. Apart from $f_\mathrm{scale}$, also the resulting standard deviations of the fitted Gaussians are given. In case of a perfect agreement between simulations and measurements, a perfect estimation and no correlation of the individual uncertainties $\sigma_\mathrm{LOPES}$ and $\sigma_\mathrm{sim}$, we would expect a standard deviation of 1. The main reason for the difference between the proton and iron simulations is that the uncertainties of $\epsilon_{100}$ are slightly larger for the proton simulations.} \label{fig_spread}
\end{figure*}

\section{Results}

In this section we compare the LOPES measurements with the REAS 3.11 and CoREAS simulations. With the exception of the absolute amplitude, REAS 3.11 and CoREAS show a very similar behavior, and can both describe the measured lateral distributions. A comparison for the amplitude parameter $\epsilon_{100}$ is shown in Figs.~\ref{fig_amplitudeComparison} and \ref{fig_histEpsDev}. REAS 3.11 generally gives slightly lower amplitudes than measured -- independently whether protons or iron nuclei are used as primary particles for the simulations, but we expect a constant shift of the measured versus the simulated amplitudes of up to $\pm 35\,\%$ due to the calibration scale uncertainty of LOPES. Thus the amplitudes of the REAS 3.11 simulations are fully compatible with the LOPES measurements. However, the amplitudes simulated with CoREAS for the same events are significantly lower, and hardly compatible with the measurements, even when taking into account the calibration scale uncertainty and the energy scale 
uncertainty of KASCADE-Grande, respectively, KASCADE. This difference between REAS 3.11 and CoREAS is not understood yet. Maybe the information loss due to the histogramming of the particles for the REAS simulations is responsible for this effect. Moreover, we recognize that the agreement between LOPES and REAS 3.11 amplitudes is slightly better for small values of $\epsilon_{100}$, i.e.~for events with lower energy. The reason might be that the energy reconstruction of the original KASCADE experiment has not been developed for energies above $10^{17}\,$eV. Hence, the energy and, consequently the radio amplitude, of some high energy events might be slightly underestimated.

Moreover, we have studied whether the spread visible in Fig.~\ref{fig_amplitudeComparison} is compatible with the expectations due to the uncertainties of the individual data points. A simple $\chi^2$-test, however, would be insufficient because it would be dominated by the mismatch in the scale: since the CoREAS simulations have generally smaller amplitude parameters $\epsilon_{100}$ than REAS, they also have smaller absolute uncertainties, which would lead to a larger $\chi^2$, but does not necessarily mean a worse agreement regarding the spread. Therefore, we have first rescaled both the simulated $\epsilon_{100}$ values and their uncertainties to get rid of the scale mismatch. The $\chi^2$-spread around the shifted lines in Fig.~\ref{fig_amplitudeComparison}, is dominated by the outliers and yields reduced $\chi^2$ values between $1.7$ and $1.9$ for the different cases. This means that we have more outliers in the comparison than expected by a standard Gaussian interpretation of our uncertainties. The 
reason for these outliers will be further investigated.

Consequently, another statistical test less sensitive to outliers is applied, to judge on the compatibility of the individual deviations in the amplitude comparison (Fig.~\ref{fig_amplitudeComparison}). Fig.~\ref{fig_spread} displays histograms of the individual deviations between the measured and simulated amplitudes $\epsilon_{100}$ divided by the individual uncertainties of each event and taking into account the scale mismatch. The width (= standard deviation of a fitted Gaussian) is slightly smaller than 1. Thus, we either have slightly overestimated the individual uncertainties or there is a correlation between the uncertainties of simulations and measurements. A possible explanation is that a part of the statistical fit uncertainty does not originate from statistical uncertainties of the amplitudes in individual antennas, but from 'true' deviations around the fit, e.g., due to the weak geometrical asymmetry of the radio lateral distribution caused by the interference of the geomagnetic and the Askaryan 
effect. If these true deviations are reproduced by the simulations correctly, this would indeed lead to some degree of correlation between the fit uncertainties of the simulations and measurements. Consequently, with the exception of a few events with a very large deviation, the spread in Fig.~\ref{fig_amplitudeComparison} is compatible with our expectations, and does not express any incompatibility between simulations and measurements.

The slope parameter $\eta$ is compared on average (Fig.~\ref{fig_slopeComparison}), but not per individual event, since it depends on the atmospheric depth of the shower maximum \cite{ApelLOPES_MTD2012}, and there is no way to reproduce the (unknown) shower maximum of the measured events with the individual simulations. Experimental observation indicates that the composition in the relevant energy range is relatively heavy \cite{ApelIronKnee2011}, and that the average mass almost certainly is in-between the masses of protons and iron nuclei. Thus, we expect that the measured distribution of $\eta$ lies in-between the simulations done for the extreme cases of a pure proton and a pure iron primary composition, if the simulations are compatible with the measurements. This is the case for both the REAS 3.11 and the CoREAS simulations.

\begin{figure*}[p!]
  \centering
  \includegraphics[width=0.92\columnwidth]{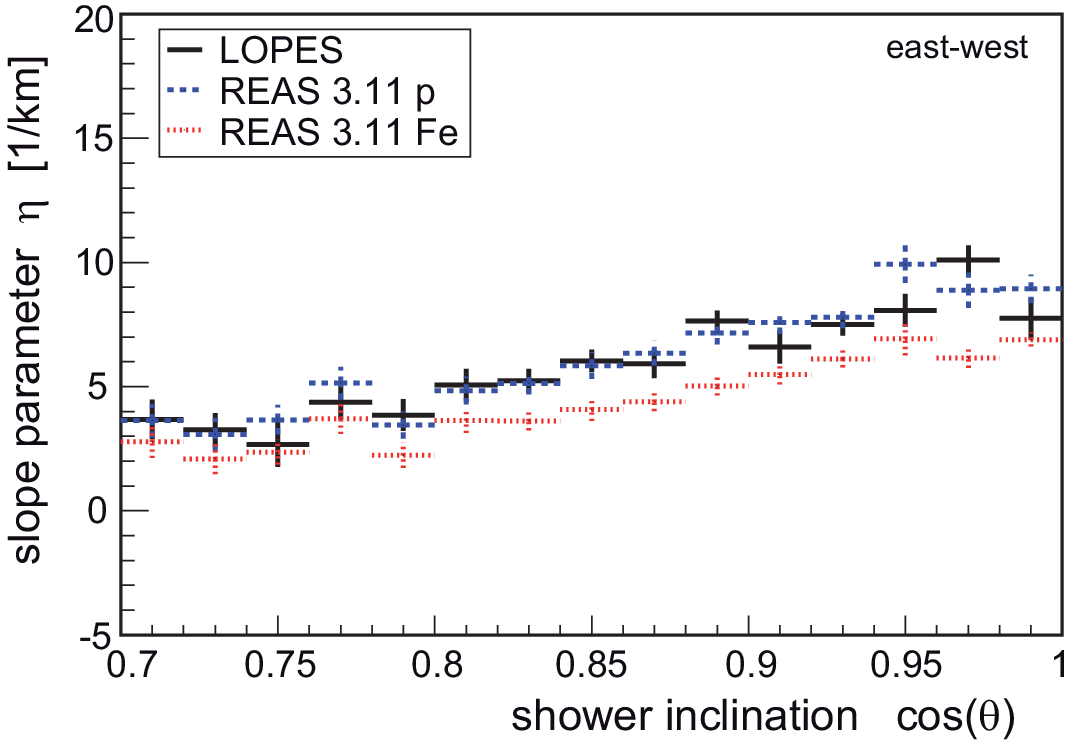}
  \hskip 0.13\columnwidth
  \includegraphics[width=0.92\columnwidth]{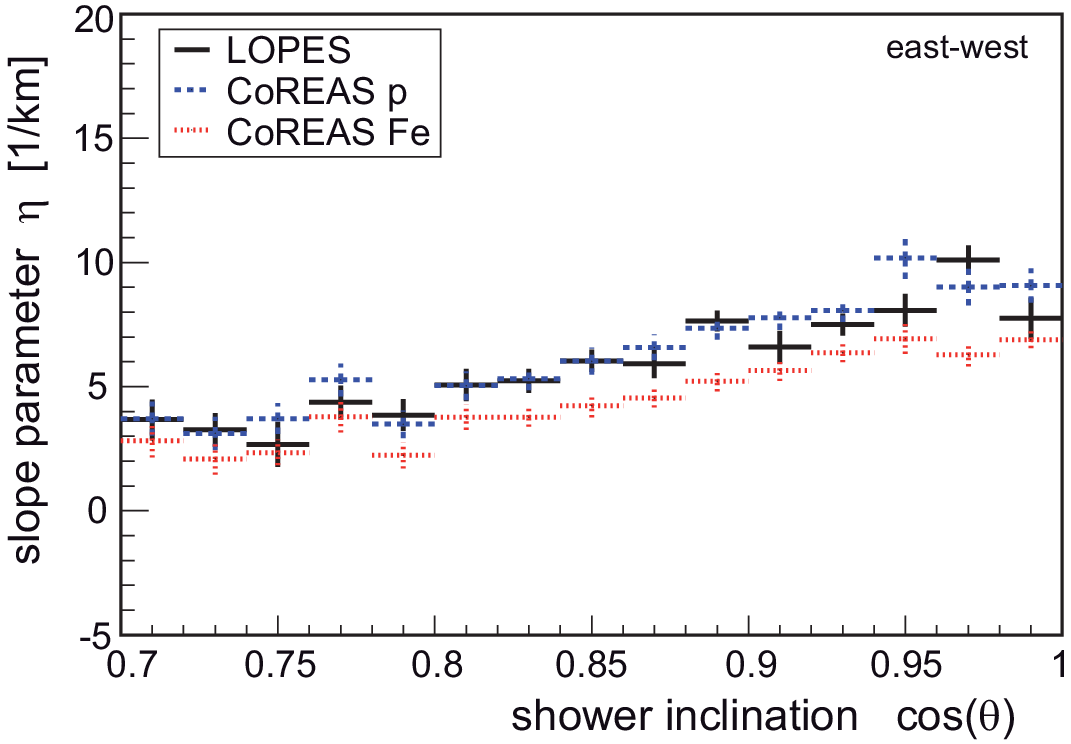}
  \vskip 0.2 cm
  \includegraphics[width=0.92\columnwidth]{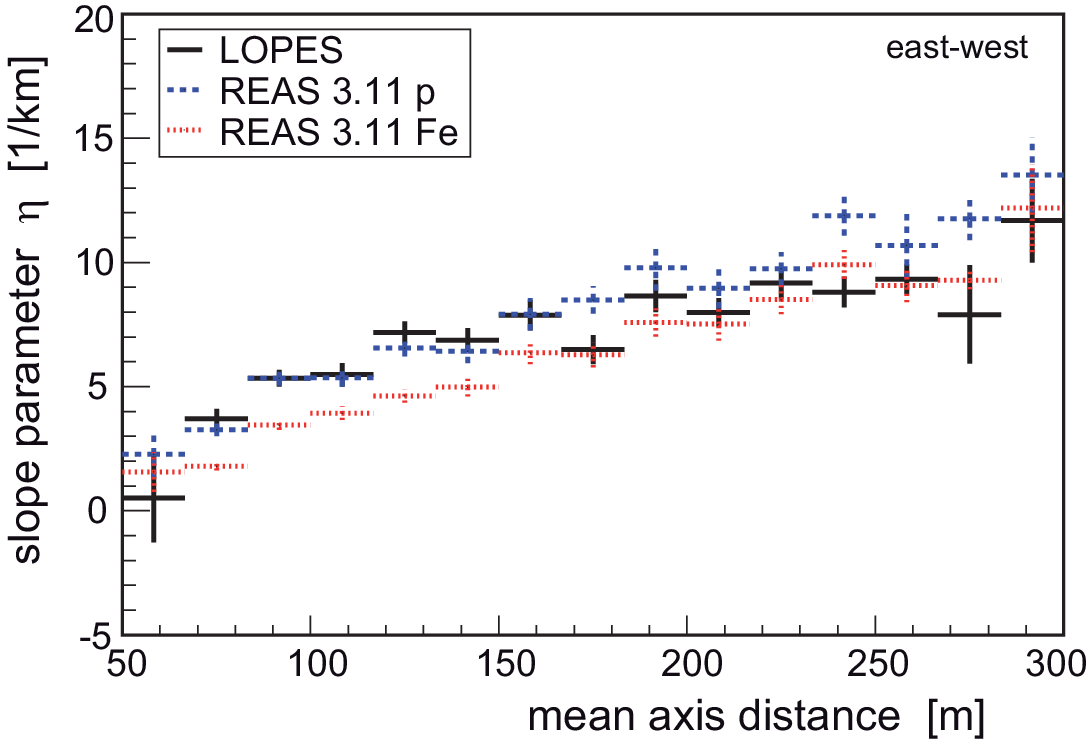}
  \hskip 0.13\columnwidth
  \includegraphics[width=0.92\columnwidth]{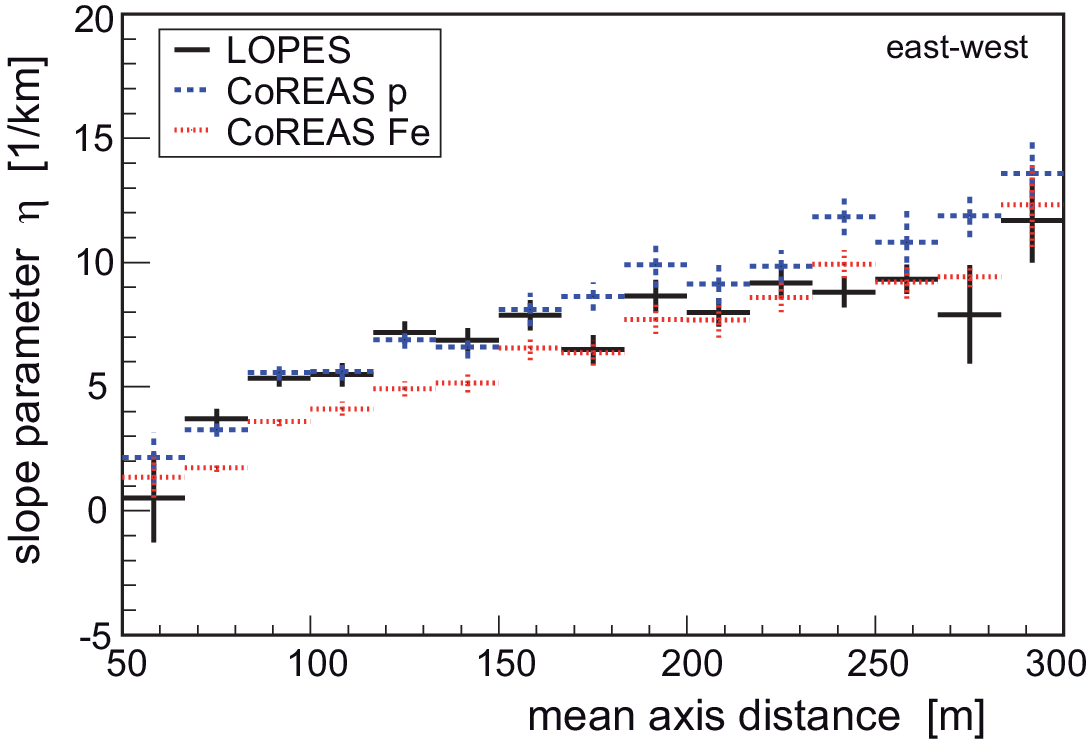}
  \caption{Profiles of the slope parameter $\eta$ of measured and simulated lateral distributions versus the zenith angle $\theta$ and the mean axis distance of the events. Although, there seems to be a slightly different dependence of $\eta$ with the mean axis distance, we cannot exclude that simulations and measurements are compatible with each other. Our event selection and reconstruction might be biased for distant events, i.e., events whose core is not contained in the LOPES array.}
   \label{fig_corEta}
\end{figure*}

The qualitative difference between proton and iron simulations is reasonable, and confirms earlier results that the slope of the lateral distribution is sensitive to the composition of the primary cosmic rays \cite{HuegeUlrichEngel2008, deVries2010, PalmieriLOPES_ARENA2012}. Thus, protons on average interact deeper in the air and have a steeper radio lateral distribution than iron nuclei. However, quantitative results for the reconstruction of the primary mass with radio lateral distributions have to be taken very carefully, since they depend on the used hadronic interaction models for the CORSIKA simulations. Moreover, the reconstruction of the composition will likely be affected by the simplified application of the detector properties for the reconstruction of the simulated radio amplitude. However, comparing the difference of the mean $\eta$ between the proton and iron simulations (approx.~$1.7\,$km\textsuperscript{-1}, Fig.~\ref{fig_slopeComparison}, and also Fig.~\ref{fig_corEta}) with the typical 
uncertainty of $\eta$ (Fig.~\ref{fig_slopeUncertainty}) provides an idea of the precision a possible mass reconstruction could have. For the LOPES measurements, which are affected by the high level of ambient radio background at Karlsruhe, the mean uncertainty of $\eta$ is in the same order as the typical difference between proton and iron induced showers. For the simulations, which do only contain numerical noise, the mean uncertainty of $\eta$ is significantly smaller than the difference between proton and iron induced showers. This shows that the primary mass can in principle be reconstructed with radio measurements, provided that the experiment is located in a region with low radio background or dimensioned for higher primary energies. 

\begin{figure*}[t!]
\centering
\includegraphics[width=0.92\columnwidth]{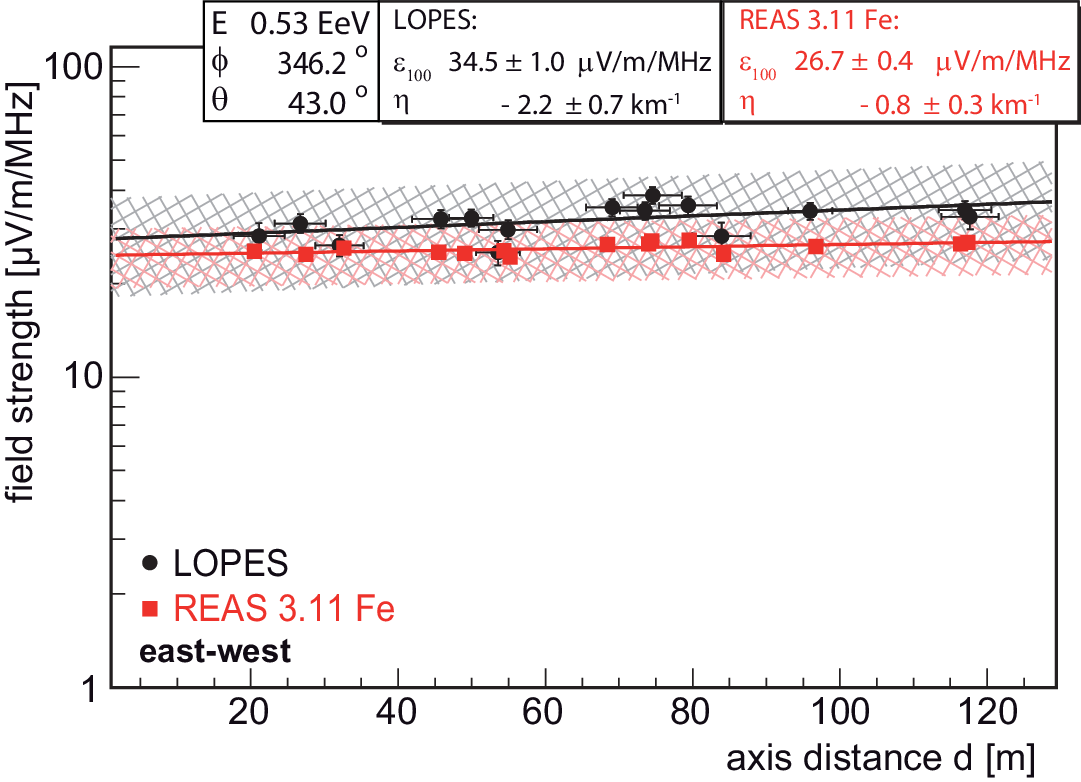}
\hskip 0.13\columnwidth
\includegraphics[width=0.92\columnwidth]{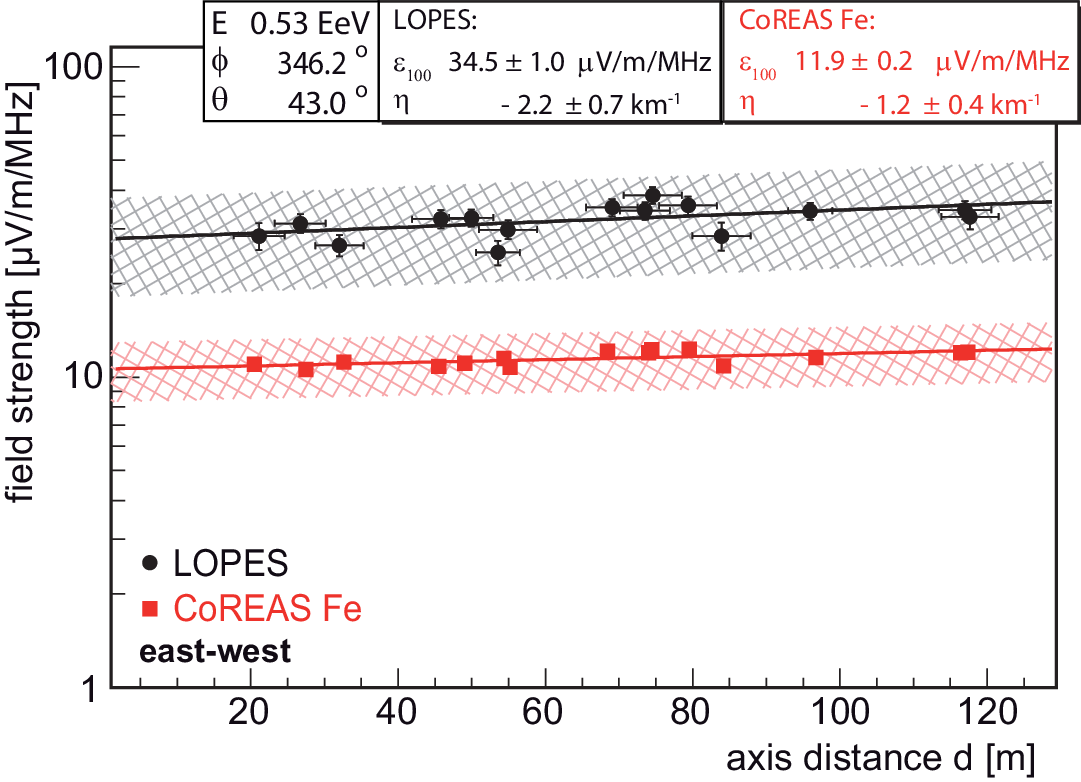}
\caption{Lateral distribution with a negative slope parameter in the LOPES measurements which in contrast to earlier versions of REAS not considering the refractive index of the atmosphere, is now reproduced by both REAS 3.11 and CoREAS.} \label{fig_risingLDF}
\end{figure*}

\begin{figure*}
  \centering
  \includegraphics[width=0.92\columnwidth]{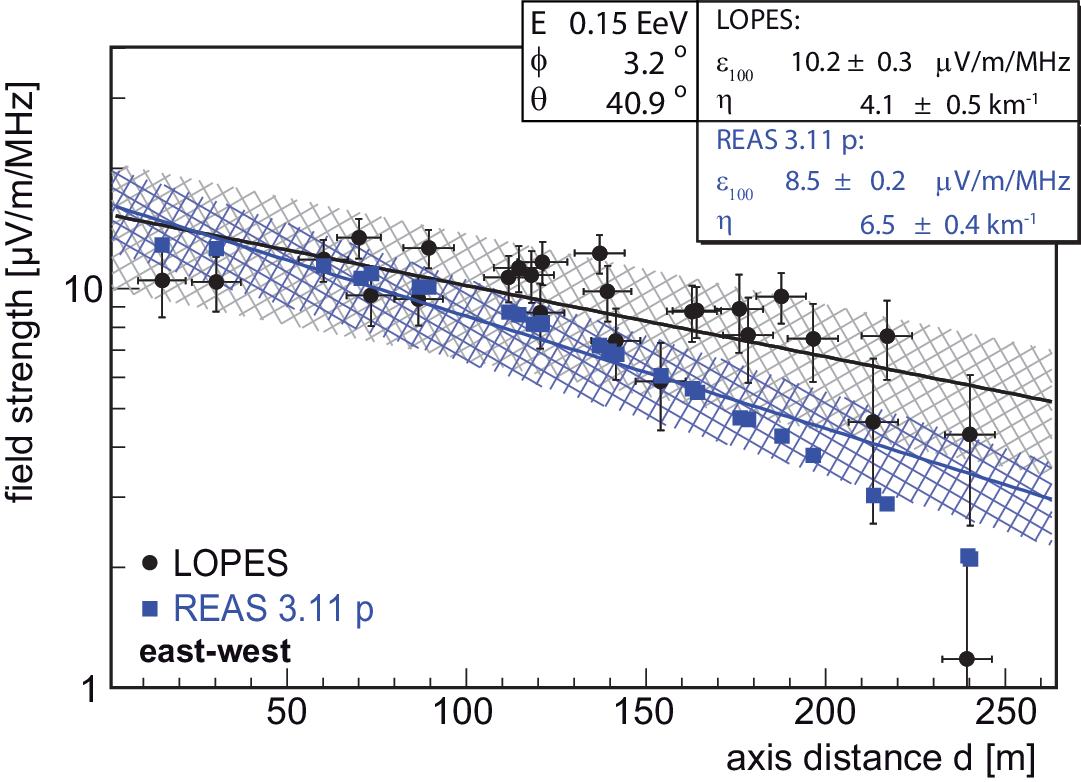}
  \hskip 0.13\columnwidth
  \includegraphics[width=0.92\columnwidth]{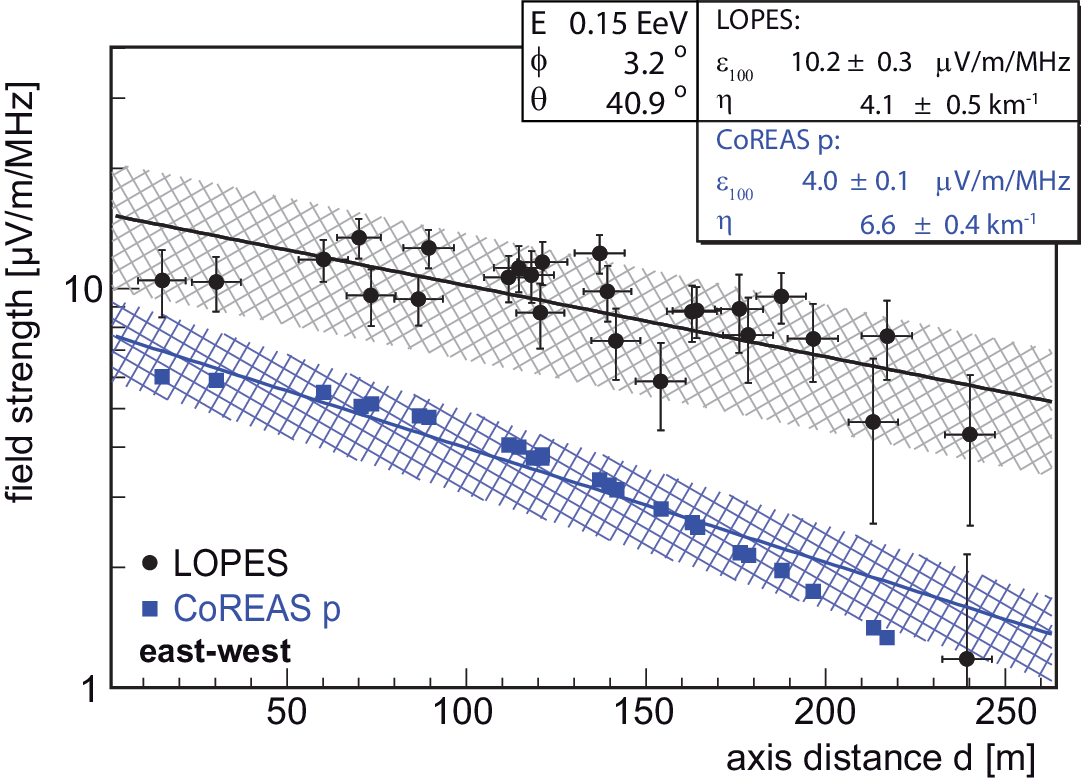}
  \caption{Example event for which a flattening towards the shower axis is observed.}
   \label{fig_flatEvents}
\end{figure*}

Also with respect to the shape of the lateral distribution both the REAS 3.11 and the CoREAS simulations are in much better agreement with the measurements than previously tested simulations which did not take into account the refractive index of the air. Already in Ref.~\cite{ApelArteagaAsch2010} we observed that the lateral distribution of some events flattens towards the shower core. With the improved fitting function, we discovered that in a few cases the slope parameter $\eta$ is even negative, i.e.~the lateral distribution rises (see Fig.~\ref{fig_risingLDF} for a rising lateral distribution, and Fig.~\ref{fig_flatEvents} for a flattening lateral distribution). First indications for rising lateral distributions have already been described in Ref. \cite{Hazen1969}, and a first theoretical explanation has been given by Ref.~\cite{AllanICRC1971}, which we now confirm. Both the flattening and the rising lateral distributions can be caused by the refractive index of the air which changes the coherence 
conditions \cite{deVries2011}. Only at the Cherenkov angle, the radio emission from all stages of the shower development arrives at approximately the same time, which results in a higher radio amplitude at an axis distance of about $120\,$m. Also in first measurements reported by LOFAR \cite{LOFAR_ICRC2011}, indications for such a bump can be seen. It can be interpreted as result of a Cherenkov-like beaming. However, one must carefully keep in mind that the main origin of the radio emission is not the polarization of the ambient medium by constantly moving charges as for normal Cherenkov light detectable at near-optical frequencies, but still the geo-magnetic deflection of electrons and positrons in the air shower, and the time-variation of the charge excess. Depending on the shower geometry, this bump can lead to rising lateral distributions -- in particular, when most of the antennas have an axis distance below $120\,$m. Concluding, both REAS 3.11 and CoREAS seem to reproduce the shape of measured lateral 
distributions correctly, at least in the frequency band and distance range of LOPES.

\begin{figure*}[p!]
  \centering
  \includegraphics[width=0.92\columnwidth]{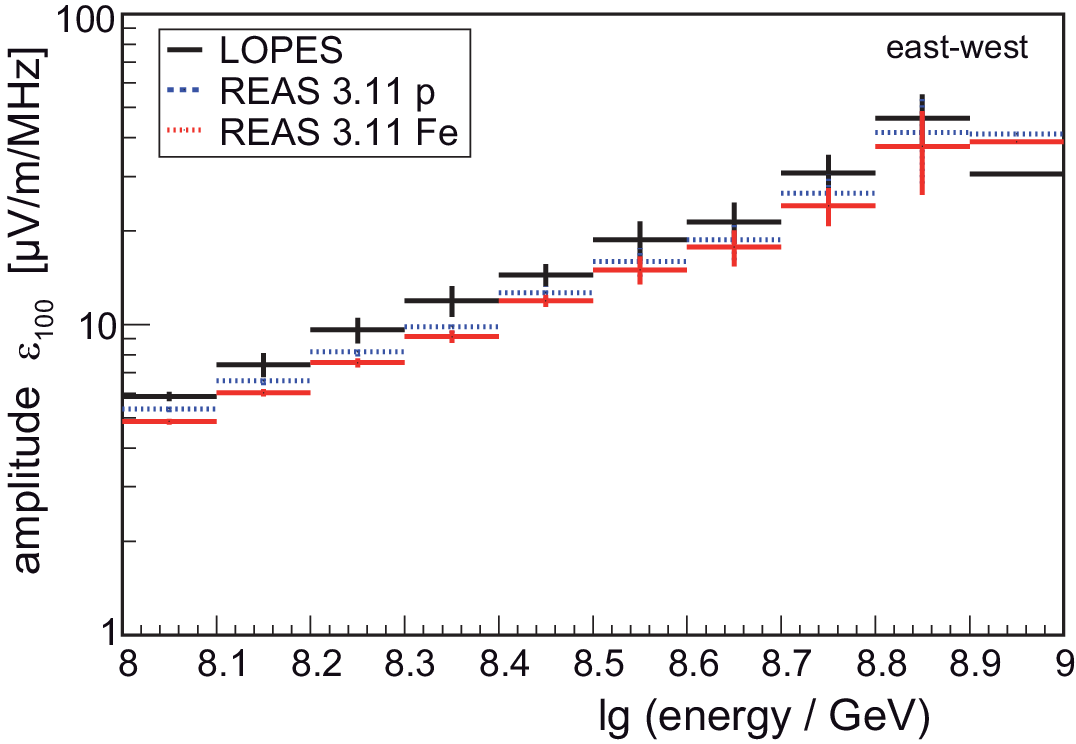}
  \hskip 0.13\columnwidth
  \includegraphics[width=0.92\columnwidth]{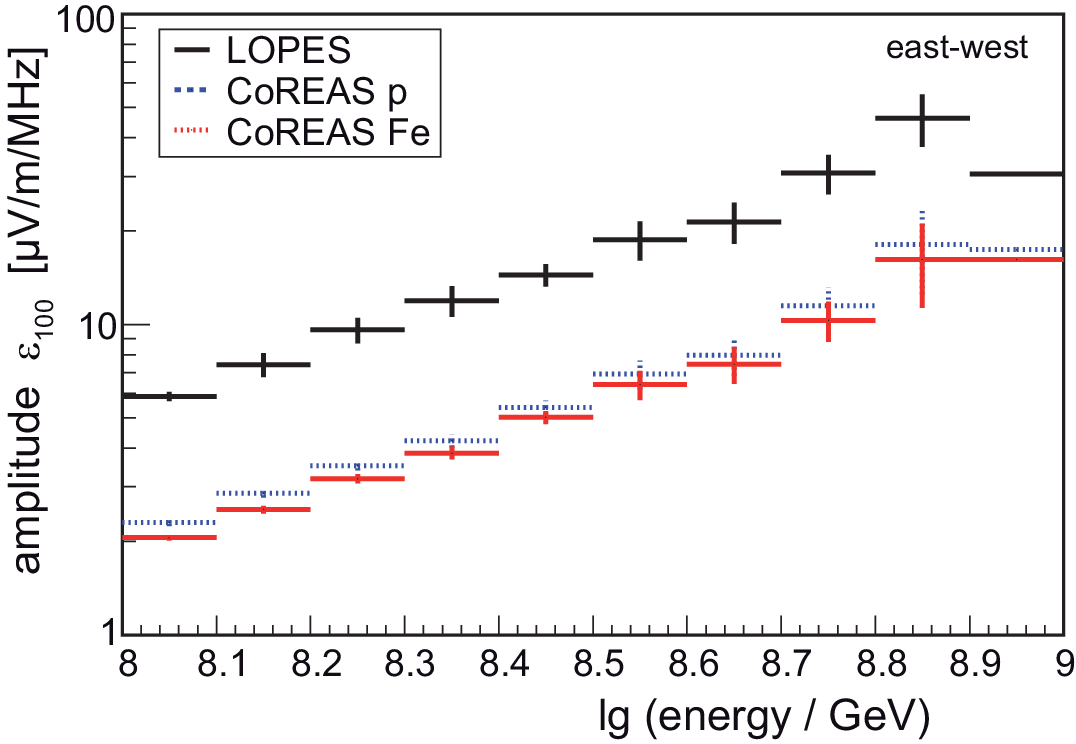}
  \caption{Profiles of the amplitude parameter $\epsilon_{100}$ of measured and simulated lateral distributions versus the primary energy $E$. The horizontal bars represent the bin width; the vertical bars denote the statistical error calculated as standard deviation divided by the square root of the number of events, where the bins without vertical bar contain only one event. While the general shift between the measurements and the REAS 3.11 simulations can easily be explained by the calibration scale uncertainty, the amplitude discrepancy between CoREAS and LOPES is significantly larger than the known scale uncertainties.}
   \label{fig_corEps}
\end{figure*}

\subsection{Correlations with shower parameters}

In the previously published analysis focusing on the experimental results \cite{ApelArteagaAsch2010}, dependencies of the amplitude and slope parameter with other shower parameters were studied. With the improved analysis and increased statistics we confirm the dependencies reported in this previous paper for the LOPES measurements. Furthermore, we have tested whether REAS 3.11 and CoREAS can reproduce the characteristics of the measurement, and find a general agreement between simulations and measurements. In particular, the simulations reproduce the dependencies on the arrival direction, which is implicitly taken into account in our study, since each simulated event is made using the geometry of the measured shower. For instance, due to the geomagnetic effect, the amplitude parameter $\epsilon_{100}$ depends both in the measurements and in the simulations in the same way on the azimuth and zenith angle.

Moreover, the amplitude parameter $\epsilon_{100}$ is clearly correlated with the primary energy (Fig.~\ref{fig_corEps}), and shows only a weak dependence on the composition of the primary cosmic rays. This confirms that the primary energy can be estimated by measuring the lateral distribution of the radio emission (and the arrival direction), as shown already in Ref.~\cite{HuegeUlrichEngel2008}, and confirmed at CODALEMA \cite{RebaiCODALEMAenergy2012} and LOPES \cite{PalmieriLOPES_ARENA2012}. Using REAS 2 and REAS 3 simulations not yet including the refractive index of the air, Refs.~\cite{HuegeUlrichEngel2008} and \cite{PalmieriLOPES_ARENA2012} give an optimal distance to measure the radio amplitude for energy reconstruction, which is the distance for which the mass of the primary particle has a minimal influence on the radio amplitude. In newer studies \cite{PalmieriICRC2013}, we have shown that this typical distance range changes only slightly to $70-100\,$m, when CoREAS simulations including the 
refractive index are used. This distance is by chance close to the typical antenna distance of $100\,$m which we use for the fit parameter $\epsilon_{100}$. In Refs. \cite{SchroederLOPES_ARENA2012, PalmieriLOPES_ARENA2012}, we demonstrated by comparing the LOPES to the KASCADE-Grande energy reconstruction that a statistical precision for the energy of better than $20\,\%$ can be achieved.

Also for the slope parameter $\eta$ we see the same correlations to other shower parameters in the LOPES measurements and in the simulations. $\eta$ is clearly correlated with the zenith angle $\theta$ and the mean distance of the antennas to the shower axis (Fig.~\ref{fig_corEta}): the more inclined the shower and the smaller the average distance from the antennas to the shower axis, the flatter is the radio lateral distribution. Both effects have to be taken into account if one aims for a reconstruction of the type and mass of the primary cosmic ray particles \cite{ApelLOPES_MTD2012}. An interesting aspect of the correlation of $\eta$ with the mean axis distance is that for near events, the measurements seem to be closer to the proton assumption, and for distant events closer to the iron assumption. However, this does not necessarily mean that the simulations would be incompatible to the measurements. Since the shower core of the distant events is outside of the LOPES array, we likely have a higher 
detection chance for events with flatter lateral distribution. Thus, the effect can probably be explained by an experimental selection bias, and consequently does not indicate a discrepancy between the simulations and the measurements. Consequently, for any analyses aiming for a reconstruction of the cosmic-ray composition, only events with shower cores contained inside of an antenna array should be used.

\section{Conclusions}

We performed a comparison of radio lateral distributions measured with LOPES and simulated on an event-by-event basis with REAS 3.11 and CoREAS, respectively. For this purpose, and in the limited distance and frequency range of LOPES, a simple, uniform exponential lateral-distribution function is sufficient. The \lq true\rq~lateral distribution is more complex and depends on the shower geometry: in many cases the lateral distribution is flattening towards the shower core and steepening towards larger distances. In some cases the lateral distribution seems to have a bump at about $120\,$m, which even can lead to a rising lateral distribution up to roughly this distance. Consequently, other antenna arrays (e.g., LOFAR \cite{NellesLOFAR_ARENA2012}, AERA \cite{MelissasAERA_ARENA2012}, Tunka-Rex \cite{SchroederTunkaRex_ARENA2012}) covering a larger bandwidth and a larger distance range in individual events might have to use more complicated lateral distribution functions.

The qualitative and quantitative results of the comparison between the simulations and measurements reflect a remarkable progress in the understanding of the radio emission generated by air-showers: In contrast to previously tested simulations, REAS 3.11 can reproduce all tested features of the lateral distribution, including the absolute amplitude, the shape, and correlations with other shower observables within the uncertainties. Thus, we conclude that REAS 3.11 can correctly reproduce the radio amplitude in the effective frequency band of LOPES ($43-74\,$MHz), and in the distance range of LOPES $d_\textrm{axis}\lesssim300\,$m.

Also CoREAS can reproduce the shape of the lateral distribution and the correlations with other shower parameters. The absolute amplitude of the radio signal predicted by CoREAS is lower by a factor of approximately two than for REAS 3.11. Thus, the absolute amplitude predicted by CoREAS is only marginally compatible with the LOPES measurements. The reason for the difference in the amplitude scale is not known yet. It is interesting since technically CoREAS is the more accurate model: it does not loose information of the CORSIKA air-shower simulation by histogramming, as REAS 3.11 does. Of course, we can not completely exclude that the mismatch is due to some kind of yet undiscovered mistake, e.g., an unknown error in either the simulations themselves, the analysis procedure, or an underestimation of the scale uncertainties, but we are not aware of any unsolved issue which could explain this difference. Perhaps, there still remains an important aspect of the air shower and its radio emission to be understood.
 In the future, we can also test other radio simulation codes against LOPES data to study this issue. Furthermore, LOFAR, a much denser antenna array operating approximately in the same frequency and distance range as LOPES, will be able to test the structure of the lateral distribution in greater detail \cite{BuitinkLOFAR_ICRC2013}. Complementary, the sparser antenna array AERA will be able to study the lateral distribution at large distances, since it features an extension of several km \cite{SchroederAERA_ICRC2013}.

With the exception of the absolute amplitude, no significant difference is observed by us between REAS 3.11 and CoREAS for the bandwidth and distance range of LOPES. However, a purely theoretical comparison between REAS 3.11 and CoREAS indicates that differences can be expected at high frequencies and very small axis distances \cite{HuegeARENAtheory2012}. Moreover, the simulated radio emission depends on the hadronic interaction models used for the underlying air-shower simulations. Recently, different hadronic interaction models converged, but still, all available models have some deficits describing air showers at ultra-high energies. Thus, a \lq correct\rq~hadronic interaction model could shift the absolute scale of the radio amplitude, for two reasons: first, the underlying energy scale of KASCADE-Grande depends on the hadronic interaction models; second, for a given energy, the radio emission depends on the electromagnetic air-shower component. In case that these systematic effects act by chance in the 
same direction as the calibration scale uncertainty of LOPES, we cannot exclude that the measured amplitudes might become compatible with the CoREAS simulations.

Both REAS 3.11 and CoREAS reproduce now rising lateral distributions which we measure in a few LOPES events. We interpret this as effect of the refractive index of air, since previous versions of REAS not including the refractive index could not reproduce this feature. Consequently, in addition to the dominant geomagnetic emission and the Askaryan effect, the refractive index of air apparently plays an important role for the radio emission of air showers. This is because the refractive index changes the coherence conditions for the radio emission which leads to a Cherenkov-like time compression of the radio signal \cite{deVries2011} and, consequently, to an amplification of the radio emission at the Cherenkov angle causing the bump mentioned above. Still, one should keep in mind one essential difference to the emission of Cherenkov light by air-showers at near-optical frequencies: While this 'standard' Cherenkov light is caused by constant charges moving with approximately constant, superluminal speed, the 
dominant origin of the radio emission is the acceleration and variation of charges in the air shower. The Cherenkov emission due to the constant movement of the charges is considered negligible at radio frequencies \cite{James2011}, and not included in the tested simulation codes.
 
Finally, we confirm with the measurements and simulations that the radio lateral distribution is sensitive to the energy and mass of the primary cosmic-ray particles. The amplitude is clearly correlated with the energy, and the slope depends on the type and mass of the primary particle, which is expected since we already know that the slope of measured lateral distributions is sensitive to the longitudinal shower development \cite{ApelLOPES_MTD2012}. At LOPES, this method seems to be limited by the high level of ambient radio background. Future experiments at locations with lower background have to show whether the precision of other established detection techniques for air showers like air-fluorescence or air-Cherenkov light detection can be achieved.

\section*{Acknowledgments}
LOPES and KASCADE-Grande have been supported by the German Federal Ministry of Education and Research. KASCADE-Grande is partly supported by the MIUR and INAF of Italy, the Polish Ministry of Science and Higher Education and by the Romanian Authority for Scientific Research UEFISCDI (PNII-IDEI grant 271/2011). This research has been supported by grant number VH-NG-413 of the Helmholtz Association.  The present study is supported by the ’Helmholtz Alliance for Astroparticle Physics - HAP’ funded by the Initiative and Networking Fund of the Helmholtz Association, Germany. The authors thank Steffen Nehls for fruitful discussions and his previous work at LOPES.


\section*{Appendix A: Results for north-south polarization}

\begin{figure*}[h!]
\centering
  \includegraphics[width=0.92\columnwidth]{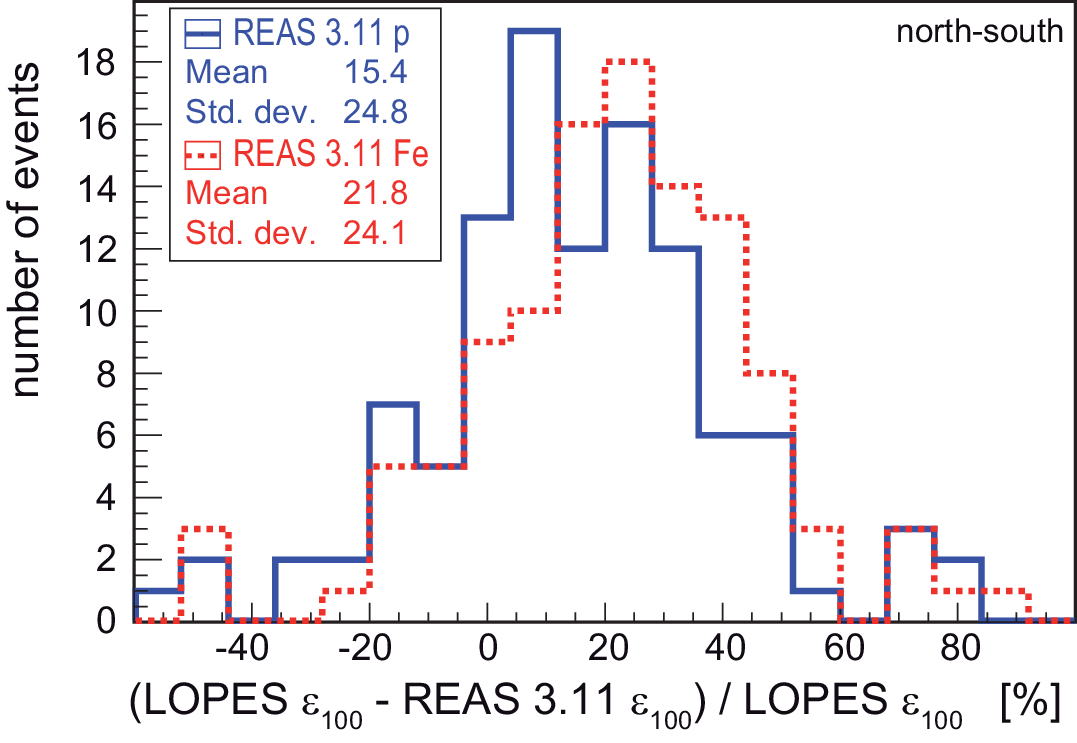}
  \hskip 0.12\columnwidth
  \includegraphics[width=0.92\columnwidth]{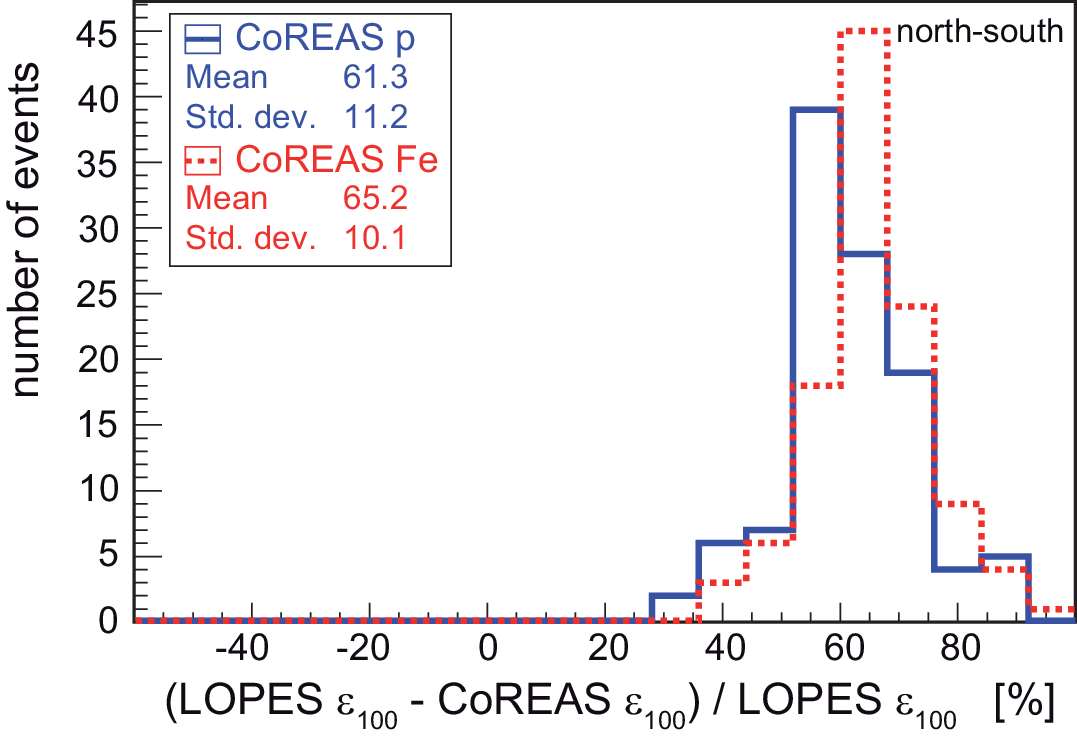}
\caption{Relative deviation of $\epsilon_{100}$ for the north-south polarization for REAS 3.11 (left) and CoREAS (right).} \label{fig_histEpsDevNS}
\end{figure*}

\begin{figure*}[h!]
\centering
  \includegraphics[width=0.92\columnwidth]{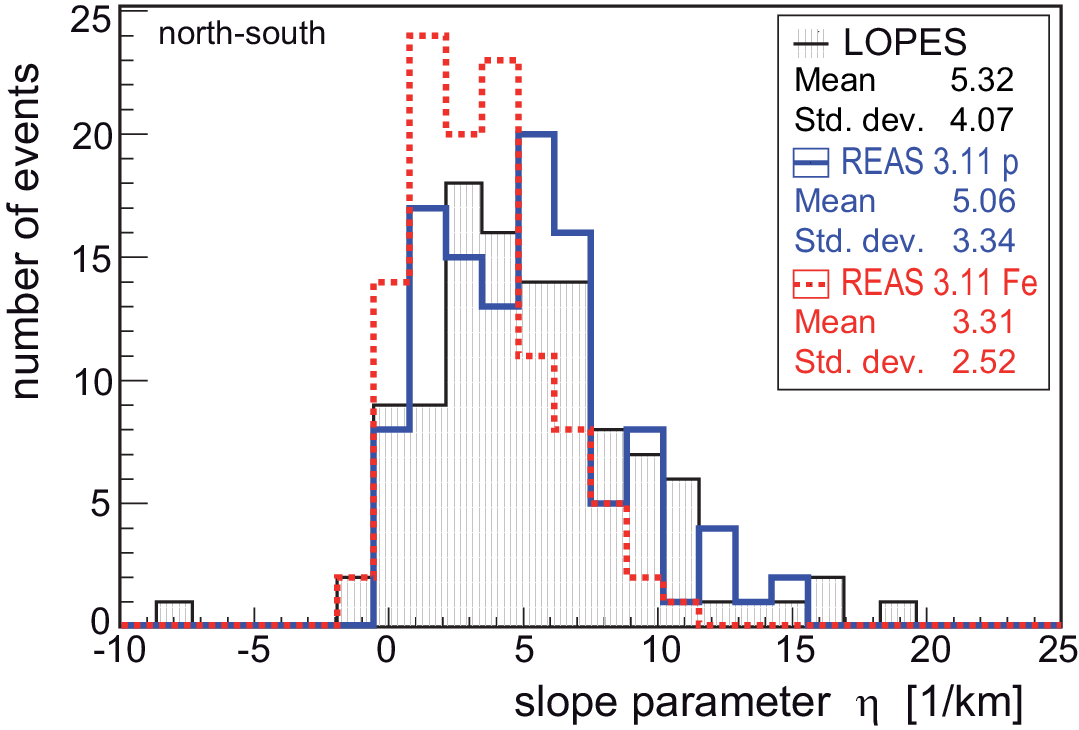}
  \hskip 0.12\columnwidth
  \includegraphics[width=0.92\columnwidth]{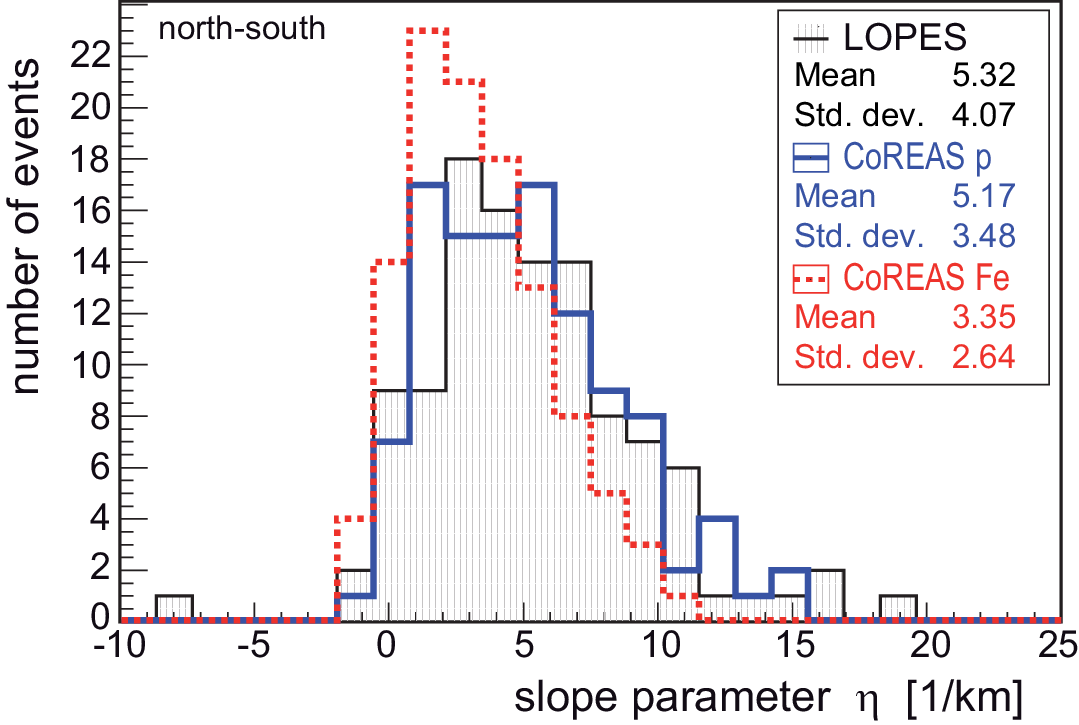}
\caption{Comparison of the slope parameter $\eta$ for the north-south polarization.} \label{fig_slopeComparisonNS}
\end{figure*}

For a part of the LOPES events also measurements with 15 north-south aligned antennas are available. Although the statistics are poorer, since only 110 events of the selection pass the quality cuts for the north-south polarization, the qualitative results are the same: The absolute amplitude $\epsilon_{100}$ at an axis distance of $100\,$m is slightly lower for REAS 3.11 than for LOPES, but compatible within the scale uncertainties of the LOPES amplitude calibration, and KASCADE-Grande energy used as input for the simulations (Fig.~\ref{fig_histEpsDevNS}). For CoREAS, $\epsilon_{100}$ is too low by more than a factor of two, and thus incompatible with the LOPES measurements. As for the east-west component, also the slope parameter $\eta$ for the north-south component is compatible with the measurements for both REAS 3.11 and CoREAS (Fig.~\ref{fig_slopeComparisonNS}).

\section*{Appendix B: Method for shower selection by muon number}
Here we give a more detailed description on the methodology we used to select a simulated shower by its muon number.

As a starting point, the air shower parameters reconstructed by KASCADE (and KASCADE-Grande) such as the primary energy, the zenith and the azimuth angle and the core position are used as input parameters for the air shower simulation. For this, a newly developed CORSIKA version (beta-version of CORSIKA 7, based on CORSIKA 6960) was used in which CONEX \cite{BergmannCONEX2007} is directly implemented \cite{PierogICRC2011}. Since CONEX uses a Monte Carlo simulation only for the first few air-shower interactions, and then makes use of cascade-equations to calculate the air-shower, CONEX saves significant computing time compared to the Monte-Carlo simulation by CORSIKA. This is true even though for the CORSIKA simulations we use the thinning technique \cite{HeckKnappCapdevielle1998} ($10^{-6}$ thinning with optimized weight limitation), to allow for reasonable computing times (about 1 day per event on a standard personal computer at the time of writing).

We have made 200 CONEX simulations for each event if the primary particle is a proton and 100 in case of an iron nucleus as primary particle due to the smaller shower-to-shower fluctuations. Input parameters for the CONEX simulation are chosen according to the KASCADE setup: observer height is $110\,$m, the cut-off energies of the muons is set equal to the energy threshold for muons of KASCADE, i.e.~muons with energies below $230\,$MeV are no more tracked in the CONEX simulation. This is important, since KASCADE is only sensitive to muons with energies larger than $230\,$MeV and the number of muons simulated with CONEX has to be comparable with the measurement\footnote{For the CORSIKA simulation, however, the energy cuts are set lower, since these muons can produce electrons and positrons contributing to the radio signal of the air shower. With CONEX implemented in CORSIKA, it is possible to choose different energy cuts for the cascade equations than for the full Monte Carlo simulation.}.

To compare the number of muons at the observer height simulated with CONEX in CORSIKA to the number of muons measured by KASCADE, we have to take into account that the used hadronic interaction model (QGSJetII.O3) generally fails to reproduce the muons of air showers above $10^{17}\,$eV \cite{ArteageISVHECRI2012, AllenICRC2011}. At the moment of writing, we are not aware of any model which would solve this issue completely. Thus, we have decided to rescale the simulated muon number such that it matches on average the measured numbers. We calculated the average simulated muon number in different zenith angle bins for proton and iron simulations separately. This way we achieved two linear, zenith-dependent correction functions, one to scale the muon number of the simulated showers with proton primaries, and one for the showers with iron primaries.

For each LOPES event, we have chosen the two simulated air showers, one with proton, one with iron primary, which have the smallest deviation from the measured muon number, after applying the correction. These showers are re-calculated with the complete Monte Carlo simulation CORSIKA (using $10^{-6}$ thinning). Since the new version of CONEX is directly implemented in CORSIKA, the same algorithms and random number generators are used, i.e. CORSIKA makes a Monte Carlo simulation of the same air-shower selected from the CONEX simulations. During this full CORSIKA simulation of the selected shower, the radio emission is directly calculated with CoREAS, as are the histograms used for REAS. Finally, the REAS 3.11 simulations are run based on these histograms.

We are aware that the way we performed the muon selection, it has only little influence on the average impact of shower-to-shower fluctuations, though the intention is to choose a simulated shower as close to the observed shower as possible. For future analyses, we can imagine several improvements to get closer to this ideal situation. With the availability of LHC measurements, better hadronic models will become available, which correctly describe air-shower muons in the energy range of interest. Then, no correction functions would be necessary to adjust the muon scale. This should be true at least if we use a complete detector simulation of KASCADE-Grande to determine the simulated muon number. In addition, further properties like the slope of the muon lateral distribution, or a mass estimator of KASCADE-Grande could be taken into account. Since the testing power of LOPES at the moment is limited by the systematic scale uncertainties of the radio amplitude, we did not find it necessary to wait for all these 
improvements for the publication of the present analysis. Still, for more accurate radio measurements at future experiments, it might be worth to take such improvements into account when applying a similar selection method for simulated showers.

\end{document}